\documentclass[acmsmall]{acmart}
\usepackage{tabularx,ragged2e,array,graphicx, pdfpages}
\newcolumntype{C}{>{\Centering\arraybackslash}X} 
\def\BibTeX{{\rm B\kern-.05em{\sc i\kern-.025em b}\kern-.08emT\kern-.1667em\lower.7ex\hbox{E}\kern-.125emX}}

\setcopyright{acmlicensed}
\acmJournal{PACMHCI}
\acmYear{2019} \acmVolume{3} \acmNumber{GROUP} \acmArticle{237} \acmMonth{12} \acmPrice{15.00}\acmDOI{10.1145/3361118}

\begin{document}

%
\title[Data Scientists and Domain Experts in Scientific Collaborations]{How Data Scientists Work Together With Domain Experts in Scientific Collaborations: To Find The Right Answer Or To Ask The Right Question?}

%

\author{Yaoli Mao}
\authornote{Equal contributions from the first author Yaoli Mao, and the corresponding author Dakuo Wang:{dakuo.wang@us.ibm.com}. Part of
work was done during Yaoli's internship at IBM Research.}
\email{ym2429@tc.columbia.edu}
\affiliation{%
  \institution{Columbia University}
  \streetaddress{525 West 120th Street}
  \city{New York}
  \state{New York}
  \postcode{10027}
}
\author{Dakuo Wang}
\authornotemark[1]
\email{dakuo.wang@ibm.com}
\affiliation{%
  \institution{IBM Research}
  \streetaddress{1101 Kitchawan Road}
  \city{Yorktown Heights}
  \state{New York}
  \postcode{10598}}

\author{Michael Muller}
\affiliation{\institution{IBM Research}}
\author{Kush R. Varshney}
\affiliation{\institution{IBM Research}}
\author{Ioana Baldini}
\affiliation{\institution{IBM Research}}
\author{Casey Dugan}
\affiliation{\institution{IBM Research}}
\author{Aleksandra Mojsilovi\'c}
\affiliation{\institution{IBM Research}}

%
\renewcommand{\shortauthors}{Mao and Wang, et al.}

%
\begin{abstract}
In recent years there has been an increasing trend in which data scientists and domain experts work together to tackle complex scientific questions. However, such collaborations often face challenges. In this paper, we aim to decipher this collaboration complexity through a semi-structured interview study with 22 interviewees from teams of bio-medical scientists collaborating with data scientists. In the analysis, we adopt the Olsons' four-dimensions framework proposed in \emph{Distance Matters} to code interview transcripts. Our findings suggest that besides the glitches in the \emph{collaboration readiness, technology readiness, and coupling of work} dimensions, the tensions that exist in the \emph{common ground} building process influence the collaboration outcomes, and then persist in the actual collaboration process. In contrast to prior works' general account of building a high level of common ground, the breakdowns of \textbf{content common ground} together with the strengthen of\textbf{process common ground} in this process is more beneficial for scientific discovery.  
We discuss why that is and what the design suggestions are, and conclude the paper with future directions and limitations.
\end{abstract}

%
%


%

%

%
\maketitle

\section{Introduction}

Thanks to the advancement of Information Technology and Cloud Computing infrastructure in recent years, a huge amount of data has been generated in the scientific discovery process ~\cite{cern, datagov} and is shared more broadly ~\cite{bosSharedDatabasesCommunities2007}. For example, the European Organization for Nuclear Research (CERN) generated 70 petabytes of data from particle physics experiments in their Large Hadron Collider (LHC) in only 2017; and they distributed and processed the data in laboratories around the world~\cite{cern}. 
GenBank in the Human Genome Project (HGP) released 212,260,377 sequences of human genome data in February 2019~\cite{genbank}. 

Such huge and complex data collection in scientific projects has gone beyond the analytic capability of a local research team in a single expertise domain, and calls for new ways of conducting scientific research. The Open Science movement started in recent years has transformed traditional science research practices to embrace more openness and re-producibility~\cite{woelfle2011open,peng2015reproducibility}. It advocates for transparency and accessibility in knowledge, data, tools, analytic processes, and interdisciplinary collaboration in the scientific discovery process~\cite{vicente2018open}. Because of the data-centric nature, most open science projects attract data scientists to collaborate with the domain experts. In this paper, we do not make fine-grain distinctions of data workers~\cite{muller2019data}, so that we denote all these data experts who often have no prior domain knowledge as "data scientists".

Many of these interdisciplinary collaborations have shown promising progress in solving hard scientific problems. For example, Critical Assessment of Structure Prediction (CASP), a biannual competition aimed at predicting the 3D structure of proteins, has attracted tens of thousands of models submitted by approximately 100 research groups worldwide and granted the top winner to a Data Science researcher team -- Google's Deepmind's AlphaFold~\cite{CASP13}.  The success of these interdisciplinary collaborations is also appealing to Human-Computer Interaction (HCI) researchers and a few papers have been published in recent years (e.g., offline data hackathon for civic issues~\cite{hou2017hacking}, or online data challenges such as in Kaggle.com ~\cite{carpenter2011may}). 

However, besides these aforementioned success stories, there are also turbulences in these collaborations. Even in the case study reporting a successful offline data hackathon event, Hou and Wang ~\cite{hou2017hacking} described a tension between the NPOs' expectations (domain experts) and the data volunteers' expectations (data scientists), which they described as a "dual goal" dilemma. In the more general open science and cyberinfrastructure contexts, tensions and challenges are not rarely seen, which have been attributed to the interdisciplinary nature of the team ~\cite{veldenExplainingFieldDifferences2013}, related motivational factors ~\cite{spencer200818} and cultural differences ~\cite{birnholtz2013cultural}, the remote and cross-culture team structure ~\cite{lawrenceWalkingTightropeBalancing2006,luo2010towards}, the data-centric practice~\cite{rolland2013beyond}, or the lack of technology and infrastructure support ~\cite{olson2002collaboratories}. 

These tensions are not new in the Computer-Supported Cooperative Work (CSCW) field. In their landmark paper, "Distance Matters", 20 years ago ~\cite{olson2000distance} Olson and Olson developed a coherent framework to describe a collaboration to be successful or not. It has four dimensions: Common Ground, Coupling of Work, Collaboration Readiness, and Technology Readiness. Though they were primarily looking at remote, not necessarily data-centric, scientific collaborations at that time (which they referred to \textit{collaboratories}~\cite{wulf1989national}), their framework has been proven to be effective in analyzing more general collaborations beyond the "remote" settings~\cite{olson2008scientific, olson2013working, olson2016converging, jirotka2013supporting, olson2017people}.  

In this paper, we continue this line of research on analyzing interdisciplinary collaborations using the Olsons' framework. We focus on data science projects and we use the bio-medical scientific domain as a case study. Bio-medical research has been one of the most active fields to embrace the open science movement, because bio-medical projects often curate and integrate many and large data sets. Yet, the data-centric projects in this domain also experienced unique challenges, partially because human lives are at stake and mistakes in analyzing data or interpreting results could lead to catastrophic consequences.

We aim to systematically explore the unique challenges that exist in the collaborations between data scientists and bio-medical scientists. Thus, we conducted this semi-structured interview study with 22 data scientists and bio-medical scientists who are involved in various open science collaborations. We have no intention to test the applicability of the Olsons' framework; rather we use it as an analytic lens to guide our coding of the interview transcripts. Specifically, the research question is: \textbf{What are the challenges in collaborations between data scientists and domain experts (i.e., bio-medical scientists) in data-centric open science projects, along each of the four dimensions in the Olsons' framework (Common Ground, Coupling of Work, Technology Readiness and Collaboration Readiness)?}   


\section{RELATED WORK}

\subsection{The Olsons' Framework and Remote Scientific Collaborations}

Olson and Olson's framework for remote scientific collaboration ~\cite{olson2000distance} brings together four major concepts that are critical to successful distributed scientific collaborations. The first concept, the coupling of the work (or the nature of work), refers to the structure and organization of the work. Ambiguous and tightly coupled tasks require higher interdependencies among collaborators and should be modularized to the same location, than the ones in loosely coupled collaborations. The second concept, common ground ~\cite{gray1989collaborating}, refers to how much common knowledge and awareness ~\cite{dourish1992awareness} collaborators have about the task and about each other. The third concept, collaboration readiness, refers to those aspects by which collaborators are motivated and willing to collaborate with each other, trust each other, as well as align their goals together. The fourth concept, technology readiness, concerns difficulties in adopting and adapting supporting technologies to fit with collaborators' current use habits and infrastructures. 

Previous HCI studies have used this framework to examine distributed collaborations and to design features to support those collaborations in various fields. 
One exemplar research work was in the international HIV/AIDS research field ~\cite{olson2002collaboratories}. This study investigated two collaborations in South Africa with case studies, and found that successful collaborations were subject to limited collaboration readiness, imbalanced technology readiness to adopt and learn advanced tools across different geographic locations, as well as inadequate bandwidth and unstable network of infrastructure. 

A more recent study re-examined this framework in globally-distributed software development teams ~\cite{bjornDoesDistanceStill2014a}. They examined four ethnographic cases of international software development using comparative analysis to explore if distance still mattered with the rapid development of collaboration technologies and people's growing familiarity and experience with these technologies and remote work over the last decade. Their findings highlighted common ground and collaboration readiness as critical factors for data- and programming-intensive collaborations, and also indicated that collaborators in this context had much higher technology readiness, and that they preferred closely coupled work even when working remotely with each other. 

In this work, we argue that we extend the Olsons' analysis of the interdisciplinary collaboration to a new genre in data-centric open science projects. Thus, we use this framework to guide our coding of the interview data, and pay particular attention to what aspects may mismatch with the Olsons' "best practices of collaboration" suggestions (e.g., successful teams should have high level of common ground).

\subsection{Teams and Infrastructures in Data-Centric Open Science Projects}

Building upon the advanced computing tools and high-speed networks for collaboration and sharing resources of e-science (i.e., supporting collaborative research via electronic networks, also unknown as e-research, cyberinfrastructure, e-infrastructure and the Grid, etc.) ~\cite{jirotka2013supporting}, the open science initiatives advocate for open access to, communication around as well as contribution to huge amounts of data sets, analytic tools, work practices and processes ~\cite{schroeder2007research}. 

In this context, novel forms of teams and ways of collaborations have emerged over recent years, transforming mere data and resource sharing base towards ecosystem-like communities of communication, practices and contributions~\cite{bosSharedDatabasesCommunities2007}. Accordingly, teams come in small and big, highly distributed geographically and self-organize themselves across traditional disciplines in the greater research community over the time. 

One example of the new team collaboration form is the PRO-ACT database ~\cite{atassiPROACTDatabaseDesign2014}, which was initially developed to pool and integrate different data sources (clinical trials, patient records, medical and family history) relating to  Amyotrophic Lateral Sclerosis (ALS). In addition to integrating these data, PRO-ACT launched two crowdsourcing competitions to the public since 2012 utilizing its database to promote ALS computational research ~\cite{ProACT}. According to the official statistics, the 2012 competition attracted 1073 solvers from 64 countries and the 2015 one drew in 288 participants, and 80 final submissions by 32 teams from 15 countries within a period of three months. The winning best algorithms
outperformed methods designed by challenge organizers as well as predictions by ALS clinicians, leading to major research publications ~\cite{kuffner2015crowdsourced}. 

One example of new ways of working is adopting Jupyter Notebook ~\cite{kluyver2016jupyter}. It allows interactive coding, visualizations, as well as building code narratives in the same UI ~\cite{rule2018exploration}. Many extensions build on top of the Jupyter Notebook system have significantly improved the data scientists' efficiency, such as the Voyager project ~\cite{zhang2018jupyterlab_voyager} for data wrangling tasks ~\cite{kandel2011wrangler} that replicate the Trifacta ~\cite{Trifacta} capabilities. Github ~\cite{dabbishSocialCodingGitHub2012} is another popular code sharing and code version control platform. It supports various types of user access so that a user can set the data to be public or private. Many data scientists use it to host their code (often in Jupyter Notebook) and manage projects ~\cite{rule2018exploration}.

Furthermore, components of machine learning and artificial intelligence have also entered the picture, in collaboration with human experts in the research fields~\cite{gil2014amplify}. Building on open code sharing with common standards, open analytics platforms such as OpenML~\cite{vanschorenOpenMLNetworkedScience2014} help users to quickly search for relevant analytical methods and reuse previous code in the community. Data analyses can be automatically processed and annotated in dataflow pipelines from how data is loaded, pre-processed and transformed, analyzed, and thus can promote mutual learning opportunities for human experts ~\cite{patterson2017dataflow,patterson2018semantic}. Very recently, DataRobot ~\cite{datarobot}, Google ~\cite{google}, H2O ~\cite{H2O}, and IBM ~\cite{watson} each have released a new AutoML solution, which aims to automatically finish low-level simple Machine Learning tasks so that Data Scientists can save some time and focus more on the higher-level tasks.

Novel forms of teams and ways of collaborations in the open science context can bring new opportunities and challenges at various steps of the data-centric collaboration process, including retrieving, preparing, and interpreting data~\cite{muller2019data}, selecting methods for analysis~\cite{patelExaminingDifficultiesSoftware2008}, and evaluating correctness of results ~\cite{kandoganDataInsightWork2014}. Hou and Wang~\cite{hou2017hacking} studied the data science process in an offline Civic Data Hackathon event. Through observation and interview research methods, they found that the broker theory is applicable to explain the tensions of collaboration between the NPO stakeholders and the data workers. Hill and his colleagues ~\cite{hillTrialsTribulationsDevelopers2016} looked at the common collaboration barriers, such as communication challenges, between multiple stakeholders, and they found that non-expert collaborators have to treat the data science process as a black box, due to the lack of timely communication.

However, the above-mentioned studies either focus on only a subset of steps of the data-centric collaboration workflow (e.g., on the data sharing~\cite{birnholtz2003data}), or on building a system or feature for a particular data science task (e.g., for data wrangling only ~\cite{zhang2018jupyterlab_voyager}). The one that tried to provide a systematic account for the whole process failed to generalize their findings to the different forms of projects (e.g., ~\cite{hou2017hacking} only looked at small teams in a data hackathon, and their unit of analysis always consisted of data volunteers working with NPOs). Thus in this paper, we contribute a comprehensive understanding of the collaborations in data-centric open science projects. And, we cover both small and large teams in data-centric collaborations.


\subsection{Interdisciplinary Collaboration Teams}
 
Olson and Olson's framework for remote collaboration mostly addresses homogeneous teams with similar expertise or experience (i.e. software engineers, or bio-medical HIV/AIDS researchers in the examples above), without a direct focus on heterogeneous teams with diverse experts. In our context, the data-centric open science projects often consist of interdisciplinary teams with distinct expertise and roles, including data scientists as the analytics experts and bio-medical scientists as the domain content experts.   

These teams have been a research focus in various research domains including HCI (e.g.,~\cite{bietz2012improving}) and cognitive science (e.g., ~\cite{hutchins1995cockpit,gorman2002expanding, derry2014interdisciplinary}). Despite some common understandings shared within the teams, a substantial portion of the domain knowledge and task understanding are distributed among different experts within the teams~\cite{gorman2002expanding}. Team performance depends on how diverse knowledge are shared and integrated ~\cite{derry2014interdisciplinary,van2007work}. What to share and how much to share have always been a critical issue yielding mixed results. On the one hand, groups should be fully informed of different and unique perspectives in order to discover an optimal solution (and thus the more the better). Stasser and Titus ~\cite{stewart1998sampling} found that in group decision-making, even though each person has unique knowledge, group members will have the propensity to discuss already shared information rather than novel, unshared information. This is known as the "shared information bias" ~\cite{stasser1985pooling} and often prevents the group from finding the alternative solution, usually an ideal or optimal one~\cite{fiedler2006information, mesmer2009information}. On the other hand, comprehensive information sharing has pooling and exchange as well as integration cost and is inefficient. Gorman ~\cite{gorman2008scientific} argued that it does not require each individual to become fully known to each others' expertise domain, but they only need to share a language enough to facilitate and evaluate team work. 

In this section, we review related theories and research that addresses sharing and integrating diversities in interdisciplinary teams. We start with the third space theory that advocates pooling different perspectives in a separate common zone, and move on to the common ground theory that supports integration and management of differences.

\subsubsection{Third Space and Hybridity}

When collaborators from different disciplines work with each other, there often a "boundary" between the two disciplines or communities. HCI researchers have proposed various theories to explain this phenomenon and these theories have guided the system design in supporting it. One notable theory that fits our context the most is the "third space" that exists "at the boundary of two disciplines"~\cite{thackara2000edge, muller2010participatory}. 

Note that this concept is different from the "third place" concept in ~\cite{oldenburg1982third}. It emerges from Bhabha's critique of colonialism, where he described that a zone of "hybridity" between two distinct cultures often came into existence spontaneously ~\cite{bhabha1994}. If each distinct culture was a "space," then the zone of hybridity, combining attributes of each culture, became something new, a "third space" that separated but also mixed those cultures.

Warr \cite{warr2006situated} extended this notion into interaction between different disciplines, suggesting preserving the situated nature of each participant's own world while creating a common space for resolving differences. Muller and Druin ~\cite{muller2010participatory} advocated the deliberate construction of a third space as part of the democratic agenda of participatory design. According to them, a third space is usually not "owned" by anyone, and subsequently diverse voices can speak and be heard in such a hybrid environment, where people can compare, negotiate, and integrate goals, perspectives and vocabularies, as well as discuss shared meanings and protocols. In line with this notion, they argued that in addition to building common ground across disciplines, differences should be adequately examined, \textit{"the mutual validation of diverse perspectives"}, and become mutual learning opportunities~\cite{bodker1988computer}. 

Within HCI, this concept of "hybridity" has been mostly used in participatory design literature, where users and designers work together across each others' disciplines to embark on a journey of negotiation, shared construction and and collective discovery. We argue that the data scientists and the bio-medical scientists in a collaboration in our context also construct a third space. As such, we expect that their behavior and their motivation in that space may differ from what they had before stepping into that zone. If so, we know various effective techniques to study and to support the collaborations in this space (e.g., spaces and places, narrative structures, games, and prototypes ~\cite{muller2010participatory}), thus we may be able to transfer these existing techniques to our context. 

\subsubsection{Common Ground: Content and Process}

With richly distributed diverse knowledge, perspectives and roles in interdisciplinary teams, common ground is required to close the gaps between differences and in turn would enable sharing and communication more efficiently~\cite{beers2006common}. This is especially important for teams of diverse experts collaborating on complex problems such as scientific research. 

Common ground originally stems from the concept of grounding in the language and communication literature ~\cite{clark1996using} and has been extensively discussed in studies of Computer-Mediated Communication (CMC) ~\cite{monk2003common}. It is defined as the sum of mutual, common, or joint knowledge, beliefs, suppositions and protocols shared between people when they are engaged in communications. And it is incrementally built on the history of joint actions between communicators. 

In CSCW, where communication becomes part of and instrumental to work activity, common ground is distinguished between two types of coordination: content and process~\cite{clark1991grounding}, which further delineates the Olsons' general notion of common ground. \textbf{Content common ground} depends on an abundant shared understanding of the subject and focus of work (know that), while \textbf{process common ground} depends on a shared understanding as well as a continual updating of the rules, procedures, timing and manner by which the interaction will be conducted (know how).  

Convertino and his colleagues studied the development of both types of common ground in an emergency management planning task that involved small teams of diverse experts. Their findings indicated that process common ground increased over time with decreasing information query or strategy discussions about how to organize activities, and in contrast, content common ground is created and tested through concept clarification and revision ~\cite{convertino2008articulating}. Furthermore, to coordinate multiple roles within teams, they suggests that a multiple-view approach, which differentiates a shared team view from role-specific details, enables teams to filter out detailed differences, construct team strategies, and allows serendipitous learning about knowledge and expertise within the team ~\cite{convertino2005multiple}, which lends support to our previous account of the third space in interdisciplinary teams.

In our interdisciplinary teams, there is a natural distinction of content domain expertise (i.e., bio-medical experts), and analytics process expertise (i.e., data scientists) when they come into collaborations with each other. We argue that the delineation of content and process common ground exists in these bio-medical research collaborations. Moreover, they may differ in what contains in content and process common ground from aforementioned communication and emergency management scenarios, which usually have a better-defined shared purpose and sometimes shared conventions and procedures as well. Additionally, over the time course, content and process common ground will also develop in different ways by both parties within teams and would need different support.

\section{METHODS}
\subsection{Participants}
Participants were recruited through snowball sampling via recruiting emails. Snowball sampling has the major advantage of efficiently locating targeted participants with adequate research expertise, who may be remote. As bio-medical scientists are not common informants in HCI studies, it is hard to find a lot of them locally. We also acknowledge the limitations of snowball sampling, such as selection bias ~\cite{atkinson2001accessing}, and we include more discussion in the Limitation section.

In total, 22 informants from 2 large enterprises (12 out of 22) and 10 research institutions (10 out of 22) in the U.S. were interviewed, reporting a variety of 26 research projects (see Table 1). Among them, 16 identified themselves with a major role of being a data scientist in the project, 6 with a role of being bio-medical scientist, and a few of them had a secondary role as a project manager or organizer. We have more data scientists due to the fact that, as participants reported, in the real-world practice, one bio-medical scientist often worked with multiple data scientists or a small domain expert panel consulted with a crowd of data scientists. The informants were quite experienced as they reported they had on average 5 years of experience in working in their expert domain (ranging from 3 years to 19 years). The projects they reported also covered a wide range of topics and team structures (from small teams with local and remote collaborators to large crowdsourcing collaborations). More details about informants and projects can be found in Table 1. Throughout this paper, data scientists will be denoted as \textbf{"DS"}, bio-medical scientists as \textbf{"BMS"}.

\subsection{Semi-structured Interview}
Semi-structured interviews were conducted during a 3-months period in the summer of 2017 as the main research method for this study, including 19 face-to-face interviews and 3 remote interviews using Skype audio chat and telephone. All the interviews were recorded and later transcribed into text. We asked the informants why they collaborated with the other domain, what data sets and tools they used, how they analyzed the data, how they communicated with each other, what outcomes they achieved. In particular, we encouraged them to recall their experience from one recent project, and we followed their storytelling with prompt questions. During the interview, informants were also asked to provide artifacts, such as source links to data sets, team meeting notes, project agendas, working documents, data analysis results and publications, presentation slides, questions and answers in community forums, and so on.

\subsection{Data Analysis and Verification}

The interview transcripts were first segmented into four dimensions of Olson's framework (common ground, coupling of work, collaboration readiness, and technology readiness) as well as specified on content versus process sub-dimensions in the common ground dimension using a deductive coding approach~\cite{crabtree1999doing}. And then for each dimension, an inductive coding ~\cite{boyatzis1998transforming} was conducted to discover salient themes regarding data, tools, processes and people. Two coders iteratively coded the transcripts and discussed descriptive memos about emerging themes from the data, and then developed axial codes that captured relationships within and across dimensions. New codes were added when necessary until theoretical saturation ~\cite{creswell2017research}. In the end, the two coders cross-checked and compared their codes. If there was a disagreement, they revisited and discussed the theoretical framework and transcripts, and then made decisions about whether to keep the codes or disapprove and toss them out.  

\includepdf[pages=-]{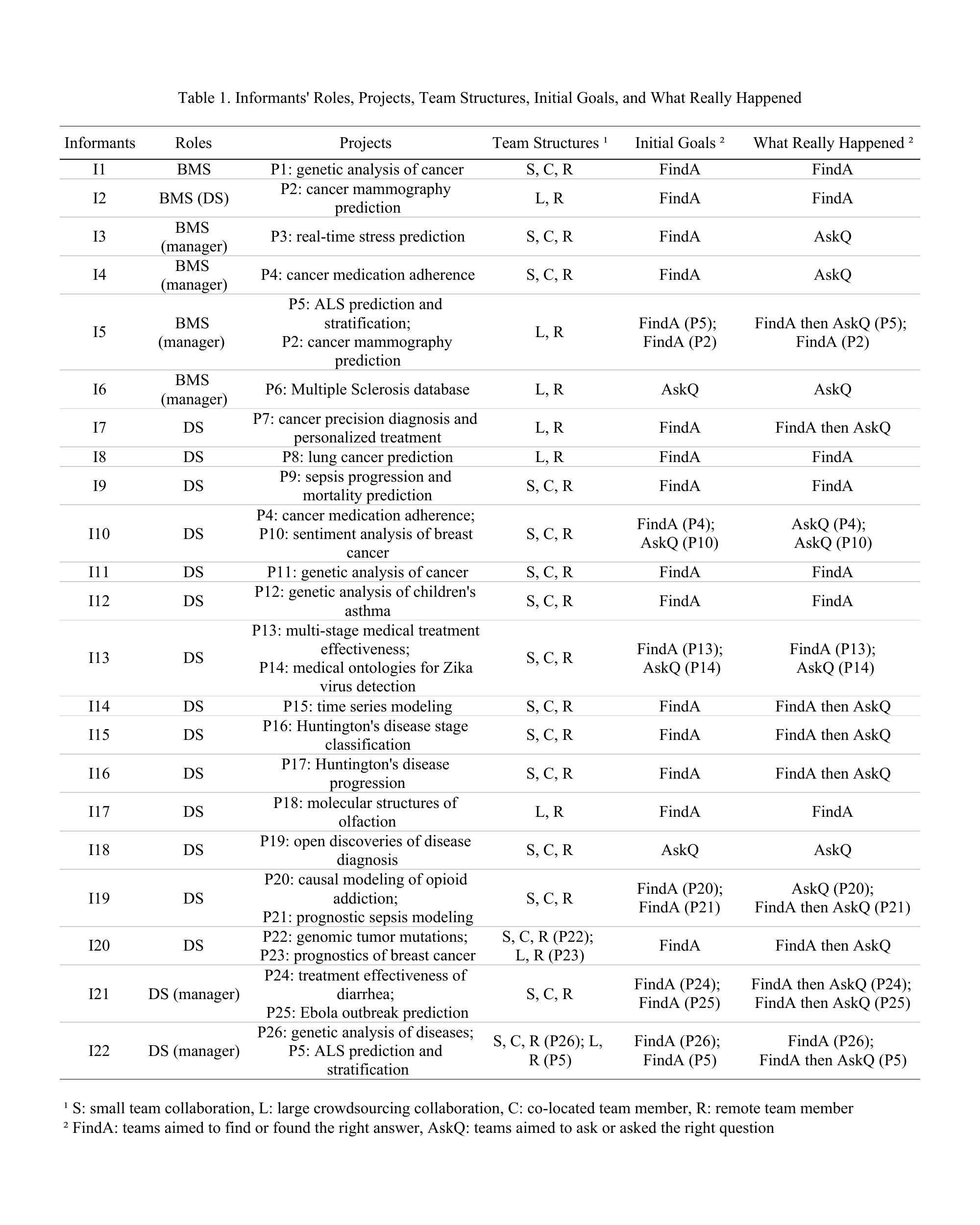}

\section{FINDINGS}
Guided by the Olsons' framework, we organize the findings in the following order: coupling of work, collaboration readiness, technology readiness, and common ground.

\subsection{The Coupling of Work}
The coupling of work, as introduced in the related work section, is often related to the nature of the project topic. The projects reported by informants cover a wide range of topics from the fundamental scientific research such as exploring the cause of disease with cell or animal experiments, to the translational and applied research that aimed to develop new diagnostics, treatments, and other related applications.

\subsubsection {Common Workflow.} 
\begin{figure}[h]
  \centering
  \includegraphics[width=1.05\linewidth]{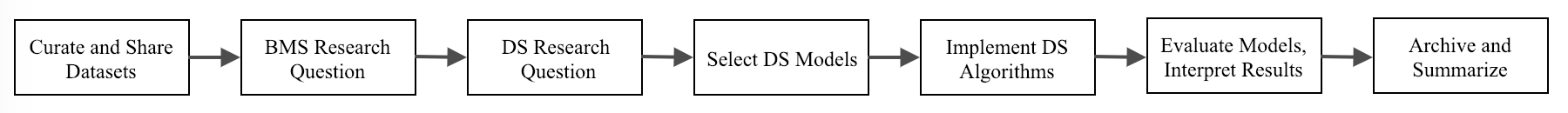}
  \caption{A simplified ideal version of common workflow}
  \label{fig:one}
\end{figure}

Despite the variety of project topics, most of these reported projects follow a common high-level workflow. Figure ~\ref{fig:one} shows an ideal and trouble-free process. The bio-medical scientists collected or curated a data set, asked a research question, and discussed it with the data scientists. Then the bio-medical research question was translated into a data science question, and a solution to the latter DS question was implemented in modeling algorithms by the data scientists. There was a final evaluation step when the data scientists synced result interpretation and model evaluation with the bio-medical scientists. Apparently, this workflow of formulating bio-medical questions, translating to DS questions, implementing algorithms, and evaluating and sometimes revising the research questions is non-divisible and highly iterative (see the Common Ground section for more results).

\begin{quote}"We brainstorm together and propose in the slack channel whenever someone has some new idea to test, try different models and quickly ITERATE prototypes in experiments to see if ideas work."(I3, BMS, P3)\end{quote} 

\subsubsection {Team Structure and Coupling of Work.}In terms of the organizational structure, all the small-group teams are managed by a researcher in the team; while in large crowdsourcing collaboration projects, a management team of organizers or project managers were responsible for structuring, monitoring, consulting and managing sub-teams along the process.

The small teams often work in a closely-coupled work style and the common understanding about how to facilitate closely-coupled work is also applicable here. For example, timely communication and coordination are pointed out as essential for the success of these collaborations.  

\begin{quote}
"... we have a lot of iterations, in deep, frequent conversations...we have weekly video meetings and frequent email checkups." (I19, DS, p20)
\end{quote}

In large-scale crowdsourcing collaborations, the aforementioned management team often helps to divide the bio-medical research question into various sub-questions, so that the multiple sub-teams working in this large collaboration can each at a time focus on one problem space which is specified clear enough, and can collaborate with other sub-teams in a loosely coupled manner. Additionally, the management team also makes efforts to regulate the proper level of coupling over throughout the process to clarify questions and engage participants. 

\begin{quote}"We also track forum questions [from other sub-teams], and provide feedback to clarify if anything [is] unclear about our data or questions...[we have] as well as webinar coaching sessions and expert advisory boards to engage participants [from other sub-teams] in learning." (I22, DS, P5) \end{quote}

\subsection{Collaboration Readiness}
Collaboration readiness refers to the collaborators' willingness and engagement level in a collaboration. Informants were asked why they collaborate, how they start to collaborate, and how their involvement proceeds over the process. From their answers, we can extract and identify the commonality of motivations in each of the two stakeholder groups (BMS, DS). We are also interested in whether their motivation and the level of engagement in the project remain the same while the project proceeds. We leave the findings about the mismatch of the motivations and engagements to the Common Ground section (See Section 4.4.4).
\subsubsection{Challenges of Maintaining Motivation}

At the beginning, people are all motivated to collaborate, because reciprocal skills and resources served as "a natural attraction for collaborations" in the data-centric bio-medical projects  (I12, DS, P12). However, these motivations and engagement levels from different experts in a team are always dynamically changing over time. Informants in small teams reported the tendency that their project soon became heavily dependent on the a few core members to manage the progress and divide the work, which can be very frustrated and reduce motivation and engagement in continuing the project.

\begin{quote}"I sometimes feel others are too much dependent on me [as both project manager and domain expert]...The team can be paralyzed...stagnant without moving forward." (I4, BMS)\end{quote}

In comparison, sub-teams in large crowdsourcing collaborations do not suffer from the heavy managemental overhead thanks to the separate management teams in the short term. However, these informants reported challenges in sustaining motivation in the longer term. These projects usually last 3 to 4 months. For many informants, it is a one-time deal. These collaborations are rarely developed into the next collaboration, especially if their solution did not came out as a winners of the internal competition, or with a concrete publication as the final credit. The short life span (a few months) of collaborations in these large crowdsourcing projects is quite opposite to traditional bio-medical research project's long life cycle (years and decades). 
\begin{quote}"Only the top winners have the opportunity to collaborate on publications after the challenge ... It is difficult to navigate to find collaborators in [large-crowd] challenge as we barely know each other." (I8, DS, P8)\end{quote}

\subsubsection{Reward Attribution and Over-Competing with Other Teams}
Being the first and finding the best result, as the nature of scientific research, encourage the competition culture, which is also reported by many informants. Sometimes it prohibits collaborations to scale up, thus limits innovative scientific discoveries. For small teams, it is obvious that the researchers in one team are competing with other teams. So they do not want to share data, processes, or tools with other teams in the research community.

\begin{quote}"We are not comfortable with sharing data or analyses before publication...[even if you share,] your work will not necessarily be acknowledged." (I10, DS, P10) \end{quote}

In large crowdsourcing collaborations that involved multiple sub-teams, over-competition is also seen as a main factor prohibiting real scientific discovery. The leaderboard type of evaluation, where each sub-team could submit a solution and all the solutions are ranked using a test data set with one metric (e.g., prediction accuracy), is problematic for scientific discovery. It motivates every team to work towards a higher ranking on the leaderboard, instead of focusing on the Bio-medical scientific discovery (e.g., whether DS results are meaningful to the current BMS question), or to find new insights from the data outside the given DS question space. After all, \textbf{scientific discovery is not only about finding incremental improvements as the right answer, it is also about asking the right and sometime disruptive questions inspired by the data}. 
\begin{quote}
"everyone copies and tweaks the best solution a bit to win a little, there is very limited innovation...but full of repetitive solutions." (I5, BMS)
\end{quote}

\subsection{Technology Readiness}
Informants reported usages of various technologies in the research process, supporting both content and progress common ground. And these technologies could be categorized into: \textbf{Co-Editing systems, Communication systems, Co-Creation systems with version control, Data and code repositories, and Expertise systems} (see Table 2). Co-Editing systems include Google Docs, Google Sheets and some other online editors, which informants used to plan or moderate project progress, and to organize project descriptions or progress summaries; Communication systems such as Slack, emails, and Skype are always useful for exchanging information quickly and tracking discussion threads; Git version control systems can help with organizing the data and code, and they are often integrated with a shared Data or Code repository system; and finally the expertise system consists of domain-specific knowledge (e.g., bio-medical ontology) where the DS collaborators can learn and query.

The challenges with teams' technology readiness are intertwined with the collaborator's backgrounds (being a DS or BMS), and are dynamically changing over time. 

\begin{figure}[h]
  \label{table:tools}
  \centering
  \includegraphics[width=\linewidth]{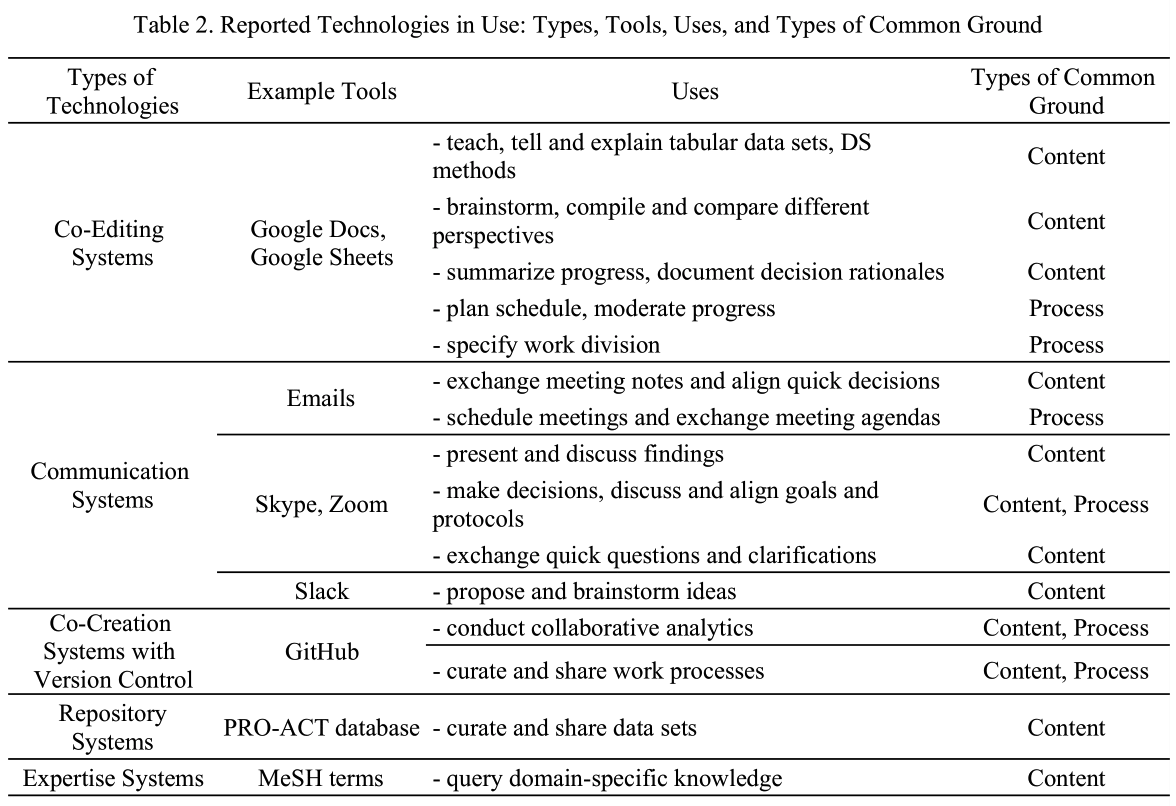}
  \label{fig:two}
\end{figure}


\subsubsection{Information Needs and Tool Preferences}

Informants in different roles reported different information needs, which resulted in different preferences over technology selections. BMSs have a focus on transparency and interpretability regarding the BMS problem, the data, the general process and the results, whereas DSs prioritize the performance, generalizability and efficiency of the DS model.
\begin{quote}
"It would be helpful to see a written documentation of pre-processing and if any transformations, any alternative methods considered or compared...These decision points can be seen clearly...lead to a trustworthy result interpretation." (I4, BMS, P4)\end{quote}
\begin{quote}"I would like to search for previous examples with similar data structures more efficiently... I also hope to extend my model built on the asthma data set as a recipe to cancer and other disease domains" (I12, DS, P12)\end{quote} 

 
Secondly, the informant's personal habits and social norms from their respective backgrounds also lead to different tool preferences. When the two backgrounds work together, they tend to find the overlapping tools that both parties can handle. This often results in the team selecting the most familiar tools that all members are comfortable with, rather than trying out more advanced new tools. When asked why the commonly used DS technologies, such as Jupyter notebooks and other cloud platforms, were not used in the team, informants explained that "persuasion cost is high"(I10, DS), and "training takes time"(I3, BMS). One BMS informant (I5) who also serves as an organizer role in a large crowdsourcing project, reported that he once tried to unify the selection of programming tools for all the sub-teams (a particular version of Python and a runtime environment) and that decision significantly reduced the sub-teams engagement and outcome.
 
\begin{quote}"We specified everyone to use python and provide written documentation using specified format in one challenge... but participation rate was much lower compared to previous challenges...And we never ask to use a unified tool again." (I5, BMS, P5).\end{quote}

\subsubsection{Fragmented Information}

Informants struggled a lot with the fragmented information all over the different systems and tools in a research project, especially in small-group collaborations where there is not a specialized management role in tracking and synthesizing information from tools used for different purposes and at different stages. This becomes more difficult when two types of common ground are managed sometimes using the same tools while at other times different tools over the process. 

\begin{quote}
"we conduct analysis on local computers using our preferred coding tools and languages, use google docs to summarize project progress internally, present slides to share progress with other stakeholders, shoot quick thoughts to each other in emails or slack messages ..."(I9, DS, P9)
\end{quote}

\subsection{Common Ground}
Most informants reported that the major challenge in their collaboration was establishing and maintaining the common ground at the beginning of the project and maintaining it throughout the process.
\subsubsection{Formulating the Initial BMS and DS Research Questions}
The common ground in formulating research questions at the very beginning refers to that BMSs and DSs work together to define a bio-medical domain-specific research question, and transform it into a computable DS question.

In small teams, it is less challenging than in large crowdsourcing projects, as aforementioned that the collaborators in small teams are quite motivated to work together at the beginning. The two research questions (BMS one and DS one) converge. The BMSs believe that they want to find an answer to the BMS question, and the DSs believe that they interpret what BMSs want into a DS question and their job is to find an answer to the DS question. 

\begin{quote}"We are working on classification of disease progressive stages...We understand from our [BMS] collaborators ... many rare disease lack proper measurement metrics to [indicate whether it is] cured or improved, and thus we have to see how clusters emerge from the data [before building classification model]." (I15, DS, P16)\end{quote}

In large-crowdsourcing projects, it is more complicated with greater ground to align. Often an expert panel consisting of organizers, BMSs and DSs is assembled to propose problems that are meaningful and impactful to BMS, as well as feasible and time-wise manageable for DS. Sometimes this expert panel even needs to have dry runs in which they simulate a team to work on this project to confirm that the question is resolvable within a period of time. I5(BMS) and I22(DS) have served in such a panel, they reported that planning such a large-crowdsourcing collaborations could take months.

\begin{quote}"In question formulation, we involve different disciplines to ask proper questions. We consult a pool of experts to ensure the problem is important and feasible as well as clear to operate on." (I5, BMS)\end{quote}

\begin{quote}"when designing a data challenge, we would arrange a dry run internally, with 1 or 2 people proposing and running 2 or 3 algorithms individually, this serves as a baseline for participants" (I22, DS)\end{quote}

\subsubsection{New Research Questions Emerge During the Project Process}
It may not be a surprise to the readers that the scientific research questions keep evolving quickly along with the project progress, but it is definitely a surprise and frustration to some of our informants. Many of them reported starting with one particular question and ended up with "a set of totally different questions" (I19, DS, P20), or sometimes "better questions"(I3, BMS, P3). In small-group collaborations, new questions emerge more frequently throughout the project process while in large-crowd collaborations, new questions often emerge at the end that point to future research directions. 

\begin{quote}
"our question evolves from what is addiction, to a set of very different questions like what is overdose, to what is abuse, to what is dependence? [This] depends on the ground truth we actually have from the data ... we later decide to focus on morphine and hypothesize about differences between natural versus herbal ones and synthesized." (I19, DS, P20) \end{quote}

Sometimes the evolved research question is a better question, and the "\textbf{right question}"(I3, BMS, P3) to ask when compared to the original one. Thus, finding an answer to the original question is less important.

\begin{quote}"we started out to ask what is stress, which context causes stress, how to measure stress ... over time we decided to focus on disease-related stress and how to build applications to monitor and design interventions...a much better question...more impactful.)"(I3, BMS, P3)\end{quote}

Overall, such evolution of questions breaks the initial common ground and requires dynamically building the new common ground. In the earlier stage, the BMSs thought they want to find the right answer and the DSs agree to find the right answer to the initial DS question. Then as the project unfolds, the BMSs may or may not realize that their true interest has changed from finding the right answer to finding the right question by asking more possible questions. Sometimes if this change is not clearly expressed, the common ground is broken. Low level of common ground, though to some extent good for allowing scientific discoveries to evolve over time, causes confusion in the team.

In I9 (DS)'s case, the initial problem raised by their BMS colleague was to design and arrange treatment resources for sepsis patients and this was transformed into a DS problem that predicts patients' life span before mortality. As DSs worked on defining mortality and dealing with missing information in the data set, BMSs came up with more questions regarding types and progression of disease severity which allow them to focus on understanding patients at different stages with various symptoms. As the poor DS commented, \begin{quote}"We were lost in which model to build and which outcome we should focus on..."(I9, DS, P9) \end{quote}

\subsubsection {Obscure Data}  
Two reasons are stated as the cause of such evolution of research questions, Obscure Data and BMS's intention to "Ask the Right Question". The open science context provided much easier access to the raw data curated and collected by other researchers in the community, but did not necessarily guarantee easy understanding of the data. Ambiguity, bias and potential missing information in bio-medical variables are particularly troublesome. Contextual information like medical practices, clinical trial routines, regulations and direct impacts on patients does not come with the meta-data or protocols but are essential for making sense of the data and asking the right question. It is a critical issue for both small-group and large-crowdsourcing projects.

\begin{quote} "I have to check across a lot of sources to clarify the implications and rule out ambiguity and biases, including standard diagnosis codes like ICD-9, pharmacy diagnosis, enrollment insurance types, typical patient demographics specific to the disease." (I4, BMS, P4)\end{quote} 

DSs reported the importance for BMSs to communicate the "data structure" with them in understanding features and relationships in the data sets. But there lacks a consistent definition and understanding of the data structure and a common language to communicate and discuss it. 

\begin{quote}"[Such information] is a hidden knowledge, a sense, and mostly gained from experience and becomes your routine" (I7, DS, P7). \end{quote}

From project to project, data structures appear in different forms, jargons and routines and are a composite concept of experiential knowledge containing:
\begin{quote} "data types and distributions like if cross-sectional or longitudinal or matrix and if there's seasonality or skew; whether there is a clear binary or continuous outcome for analysis or it is high dimensional multivariate data" (I13, DS, P13)\end{quote}

Difficulties in communicating data structures could lead to further challenges in evaluating the methods and interpreting the results, and cause BMSs' frustration and distrust around this "big black box", as quoted from I4 (BMS, P4) and I17 (DS, P18).

\subsubsection{Ask The Right Question} 
What are alternative ways to ask questions? BMS informants often reported their intentions to ask the right question by asking more alternative research questions besides the initial one. They are also frustrated that they do not know if the translated DS question is a good one or not. DSs are trained to abstract and simplify a realistic problem into a analyzable and computable one, thus BMS Problems are more often translated into Prediction problems, in which an outcome is well-defined, and the model and algorithm is "mature and well developed", and the evaluation is standardized by "a mathematical loss function" (I20, DS, P22\&P23).

\begin{quote}"In our bio-medical training, alternative hypotheses are important ways to conduct research. I conducted a lot of literature review to understand what has been established and what is the gap in reasoning. However, when translating a bio-medical question into a data science one, I often wonder what are alternatives. The process seems to be very intransparent."(I4, BMS, P4)
\end{quote}

DSs' prone-to-predict tendency could be explained by both different interests and evaluation criteria valued and rewarded by BMS and DS fields. And it adds to the misalignments in the common ground as DSs are partially instrumental to BMS. BMSs are mostly interested in the results which are meaningful for interpretations and useful for interventions; DSs are driven by developing competitive, innovative and sophisticated methods such as \textit{"no one has tried before" (I12, DS, P12), "beat existing methods in accuracy" (I8, DS, P8), "complex mathematical models"} (I15, DS). 

\begin{quote}"discovery [instead of prediction] that can be useful to provide actionable insights for high-stake life or death issues ... we are always reproducing predictive models with higher predictive capabilities in the field. However, bio-medical problems rarely have a clear outcome to make predictions... we are more interested in what intervention can be done rather than whether a prediction is accurate." (I4, BMS, P4)\end{quote}

In small project teams, this prone-to-predict tendency seems more severe; while in large-crowdsourcing collaborations, wisdom of the crowd is able to pool diverse perspectives and considerations to look at the same problem. 

\begin{quote}"At later stage of the data challenge, an ensemble method, which is a linear combination, was applied to aggregate across the winning teams' individual models, to learn from different focuses and merits in different approaches and for discovering new insight." (I20, DS, P23)\end{quote}

\section{DISCUSSION}
\subsection{Successful Collaborations In Scientific Discovery}

These reported open science projects are characteristic of their team sizes (small or big scale), distinct complementary interdisciplinary nature (bio-medical as the content domain and data science as the solution domain), tight collaboration process (rather than simply resource sharing), as well as the long-term and transformational nature of scientific discovery. Using the Olsons' four-dimensions  framework for successful distributed collaborations in scientific research (coupling of work, common ground, collaboration readiness, and technology readiness), we organize our results according to this framework, and focus mostly on the common ground dimension as the major challenge. Particularly in the diverse contexts of data-centric open science projects, we take into accounts the contrasts of small and big teams, and the dynamically evolving nature of scientific discovery. 

\subsubsection{Coupling of Work, Collaboration Readiness, and Technology Readiness}

Our findings regarding the coupling of work echo what the Olsons' framework suggests: the tight coupling within small teams requires timely communication and coordination. Loose coupling was a pre-requisite for successful distributed collaborations, such as in the large-scale crowdsourcing projects. However, most of these open science projects were non-divisible and highly iterative, which made assigning modular work for each location and setting up routine impossible. Similar to the result from a previous study ~\cite{bjornDoesDistanceStill2014a}, tight coupling under proper management, was not challenged by remote technologies but rather helped to enhance common ground and collaboration readiness.    

Our findings suggest that collaboration readiness is challenging within small project teams as well as in sub-teams in large-scale projects. In the Olsons' original framework, collaboration readiness was seen as how team members were motivated to engage with each other. However, in these reported open science projects, more aspects of organizational structures came into play, including dependence between different expertise within teams, relationships between teams, as well as over the time dimension.

Similar to previous research on domain experts collaborating with computer scientists in cyberinfrastructure ~\cite{lee2010driven} or in civic data hackathons~\cite{hou2017hacking}, each party comes in with a different research agenda, which is analyzed as "the dual-goal dilemma". This tension also exists between BMSs and DSs in our study, and manifests itself into the tension between asking the right question versus finding the answer in common ground. It is important to carefully weigh both sides' interests in the organizational structure of the team; otherwise either one side will become "merely" instrumental as consultants and implementers to the other~\cite{atkins2003revolutionizing}. 

We could also learn from existing successful experiences. The introduction of a broker role to serve as the bridge between domain experts and data scientists to translate one stakeholder's goals to the other proved useful in civic data hackathons~\cite{hou2017hacking} and large-scale collaborations~\cite{wenger2010communities,paepcke1996information, pawlowski2004bridging}. Thus, we expect to see a smoother and more successful collaboration if someone in the collaboration can play the broker role.

In terms of technology readiness, informants reported a wide range of tools, ranging from Co-editing systems to communication systems, and the reported use practices are consistent with prior literature (e.g., ~\cite{wang2016people, wang2019slack}) thus are not listed. At the same time, BMSs and DSs have different information needs and tool preferences and when they come into collaboration as a team, they usually choose the most familiar tools for all the members (mostly aligning with BMSs' tool comfortableness) rather than trying out new advanced tools. This is similar to prior findings on Co-editing technologies ~\cite{wang2017users}, and a National Science Foundation report warned if domain experts are weighted too heavily in the organization, procurement of existing technologies will be much overemphasized compared to development or adoption of new technologies ~\cite{atkins2003revolutionizing}. Furthermore, our informants also expressed concerns of managing multiple tools as well as trying out new tools to meet the needs of quickly-evolving common ground. In particular, tool interoperability between team members and across the research process was critical. Compared to project management in general workplaces, managing interdisciplinary research projects can be more difficult due to their ambiguous and ever-evolving nature, and to the lack of awareness and resources allocated to management~\cite{kirkman2012across}. Thus, a training of project management and new tool adoption may be helpful.  

\subsubsection{Ever-Evolving Common Ground and Better Scientific Discovery}

We found that common ground continued to be a key issue for both small and large-scale project teams in open science. In our findings, a "third space" ~\cite{muller2010participatory} naturally came into being when BMS and DS started collaboration. In this shared common space, separate from each of their own domain, BMS and DS initiated a concrete common ground of what the BMS and DS research questions are, building dialogues and terms around the "data structure" with hybrid languages and training from their distinct domains, negotiating tools shared by the entire group, as well as showing promises in constructing new understandings of the initial problem. In particular, boundaries between BMS and in this "third space" continued to blur and thus new possibilities of asking questions emerged. Many informants in our study reported the unexpected turns of their research projects, starting from one question and ending by answering another better question or coming up with more alternative questions. 

This echoes Convertino's previous findings in group emergency management~\cite{convertino2008articulating}. In both cases, process common ground regarding know-how seems to keep increasing through joint activities within the team, while content common ground keeps being re-articulated, broken and revised throughout the process. Different from teams in general workplaces that are driven efficiently towards clear business goals and specific performance evaluations that match one optimal solution to a well-specified problem~\cite{olson2000distance}, this differentiation of content and process common ground and their development and interaction with each other over the time course become more salient and critical. And in our context of bio-medical research collaboration, the content common ground is in the form of research questions encapsulating a complicated composite of variables and relationships in ambiguous data sets, and the training and nature of Bio-medical research to "Ask the Right Question" as well as research goals, in comparison to new concepts and terms in emergency management context~\cite{convertino2008articulating}. This is consistent with the account that scientific discovery teams are operating on the foundation of alternative explanations and different voices, to explore and rule out many possibilities rather than exploiting a set of existing successful solutions~\cite{sijtsma2016playing}.  

Moreover, the increasing process common ground, in fact, allows the breaking and updating of content common ground to be possible. Specifically, the need for new communication protocol around what is "the right question" is on the rise over the research process. Further effort is needed to recognize changes in both types of common ground from both BMS and DS communities. Failing to do so may cause confusion and low productivity, less ideal scientific discovery. For example, teams could get confused about what is the current content common ground without the support of increasing process common ground, get "frozen" with the established content common ground ~\cite{kruglanski1996motivated}, "seized" by shared information bias ~\cite{stasser1985pooling} and settle on "premature consensus" or "early closure" ~\cite{kerr2004group} of less optimal questions or solutions instead of advancing to the next stage of scientific discovery.

In order to examine the validity of this preliminary finding and understand detailed needs of BMS and DS, further research is necessary to devise measurements for both content and process common ground specific to bio-medical research collaboration in the wild compared to in controlled experimental settings~\cite{convertino2007does, convertino2008articulating}.

\subsection{Principles for Technology Design}
From our findings, the biggest challenge in open science projects seemed to be the quickly-evolving common ground with a purpose to advance scientific discovery by asking the right question, instead of finding answers within a constrained space. It affects the other three dimensions in the Olsons' framework: coupling of work, collaboration readiness, and technology readiness. It is also related to the theme of integration of heterogeneity from the seven common themes for designing and researching current and future e-Research cyberinfrastructures, articulated by Ribes and Lee in a theoretical summary ~\cite{ribes2010sociotechnical}. We refer to the related literature and discuss principles and potential designs to address this issue.

For both small-group and large-crowdsourcing collaborations, asking the right question depends on steadily developing progress common ground in terms of conventions and procedures, while constantly re-establishing content common ground through more and better questions as the research focus. Consistent with the "third space" in interdisciplinary collaborations, a multiple-view approach that differentiates a shared team view from role-specific details has been found to be effective for group tasks~\cite{convertino2005multiple}. In terms of what is to be shared in the common view, two principles are suggested here. Firstly, a divergent-to-convergent two-stage path ~\cite{paletz2010social} to help structure the tightly coupled communication in the "third space". This path starts from pooling and sharing different perspectives for more questions, and heads to comparing and evaluating for better questions. The communication systems reported in Table.2 may be further improved to support and keep track of this divergent-to-convergent model by explicitly enabling users to brainstorm ideas, then summarizing ideas, and later evaluating the different ideas in it. Secondly, it would be helpful to differentiate the two types of common ground as they develop differently over time and affect team members and their roles differently. For example, team members can see not only the current status of shared objects, but also the changes in historical states~\cite{greenberg1990sharing}. This would be similar to how today's co-editing systems (Table 2), integrates with version control systems~\cite{wang2015docuviz}, raising awareness of changes over time in separate views for common content knowledge and process protocols.

\subsection{Project Management Guideline}

On the other hand, a non-technical solution may be complementary to the technical ones for the small teams without a specialized project manager that face challenges in managing fragmented and repetitive information, or maintaining collaboration readiness over time. This might be due to the lack of awareness, expertise and resource allocation to project management compared to the conduct of the science research~\cite{olson2013working}. Training workshops could be helpful for researchers to learn about good team leadership, facilitation, and process management. Leveraging on existing technology, a shared vocabulary wiki page and data documentation could be helpful for the DSs and BMSs to keep in sync of the understanding and collaboration awareness, what questions the BMSs are interested in right now, and what questions the DSs are working on. Furthermore, specialized project management tools with interoperability across other tools could be developed to address such issue.

\subsection{AI as a Partner in the Future of Data-Centric Scientific Discovery}
We have seen a gap between the BMS and DS in our study in the sense of asking questions, translating the BM question into a correct DS question, and interpreting the DS results. BMSs sometimes distrust the results. And DSs sometimes have a different priority in methodologies and solutions that might over-simplify the question. More importantly, shown in our results, BMSs need an iterative loop with lots of redundant DS attempts to be inspired by the data, the models, and the results generated by DSs. 

 These differences, if not properly shared, communicated and integrated within the group, could become hidden biases that hold back the progress of scientific discovery. The work of Tversky and Kahneman ~\cite{tversky1974judgment, kahneman2011thinking} argues that people, even scientists and data scientists who are professional in analyzing data, have trouble thinking statistically and reasoning about the data. This contributes to the growing reproducibility crisis in recent years, in which results of many scientific studies are difficult or impossible to replicate in subsequent investigation~\cite{peng2015reproducibility, staddon2017scientific}. And it can have a significant impact on judgments and decisions around data and even reverse decisions. It has been a robust phenomenon in bio-medical field, affecting diagnosis, treatment and lifesaving, medical resource allocation and management~\cite{kuhberger1998influence,gamliel2010attribute, armstrong2002effect}.


In recent years we have seen a fast and vast research effort of using one special group of machine learning techniques to design another machine learning algorithm ~\cite{nargesian2017learning,liu2019formal}. In particular, AutoML (automated machine learning) refers to a type of technology that only requires users' minimal effort in uploading the data set, specifying the target and the DS method type (e.g., regression or binary classification), then the AI can automatically generate new features, select features, search alternative models and tune the models' parameters to reach an optimal solution (often quantified in accuracy metric)~\cite{khurana2016cognito}. With these systems, now the non-data-scientist users like BMSs in this paper may have the capability to directly build machine learning models with their domain-specific research questions. In a potential AI-human collaboration future, BMSs and DSs can leverage AutoML systems to quickly generate many ways to ask questions (including predictions and open discoveries) at different stages of the research process, and the machine may have less biased judgments despite the DSs' or BMSs' competing interests. AutoML may never fully liberate the human DSs, but we expect it could work as a partner in the human DS teams (e.g., as conversational agents illustrated in~\cite{shamekhi2018face}) and help the BMSs in this Right Question formulation process. Certainly it is hard to achieve because in addition to technical development, many non-technical aspects (e.g., anthropomorphism~\cite{tan2018projecting}) need to been taken into account. But, we choose to work toward this future because it is hard.
\section{LIMITATIONS}

One limitation of this study is the snowball sampling method, which might introduce selection bias ~\cite{atkinson2001accessing}. These informants within the reach of our social network might be above the average active level in participating in open science collaborations and report more positive experiences. Additionally, all our informants are based in the U.S., which do not necessarily represent diverse cultural differences and a wide range of geographical distances in open science collaborations. 

The semi-structured interview method is also limited in relying on informants' self-reports, which are subjective, single-sided and probably over-simplified. In order to understand the details of dynamic interaction between experts from different disciplines, it is important to design specific measurement for both content and process common ground, and observe contextual interaction within teams in real scenarios and conduct longitudinal case studies to track their processes along the research pipeline. 

Lastly, we picked bio-medical research as our target domain and it is yet to be studied how these challenges would vary for other domains involved in data-centric collaborations in open science, such as physics, geology, psychology. 

\section{CONCLUSION}
This work reports the challenges that emerged from scientific collaborations between data scientists and bio-medical scientists through interviewing 22 participants. 
Our study contributes to the existing literature by providing a systematic account for different stakeholders' practices in scientific collaborations. In particular, we differentiate content common ground versus process common ground as a finer-grained level of the common ground concept. We discovered that scientific collaborations require constant breaking of the content common ground while accumulating process common ground, in comparison to most decision making or problem solving scenarios, where only one decision or solution is the final product. Our results shed light on the better practices for future interdisciplinary scientific collaborations. And the system design suggestions are also valuable and actionable for developers and designers who are developing data analytic tools and cloud sharing platforms. 

\section*{Acknowledgement}
\addcontentsline{toc}{section}{Acknowledgement}
We thank all the interviewees who shared their research stories and resources. This work was conducted under the auspices of the IBM Science for Social Good initiative.


%
\bibliographystyle{ACM-Reference-Format}
\bibliography{sample-base}


\begin{thebibliography}{104}


\ifx \showCODEN    \undefined \def \showCODEN     #1{\unskip}     \fi
\ifx \showDOI      \undefined \def \showDOI       #1{#1}\fi
\ifx \showISBNx    \undefined \def \showISBNx     #1{\unskip}     \fi
\ifx \showISBNxiii \undefined \def \showISBNxiii  #1{\unskip}     \fi
\ifx \showISSN     \undefined \def \showISSN      #1{\unskip}     \fi
\ifx \showLCCN     \undefined \def \showLCCN      #1{\unskip}     \fi
\ifx \shownote     \undefined \def \shownote      #1{#1}          \fi
\ifx \showarticletitle \undefined \def \showarticletitle #1{#1}   \fi
\ifx \showURL      \undefined \def \showURL       {\relax}        \fi
\providecommand\bibfield[2]{#2}
\providecommand\bibinfo[2]{#2}
\providecommand\natexlab[1]{#1}
\providecommand\showeprint[2][]{arXiv:#2}

\bibitem[\protect\citeauthoryear{Armstrong, Schwartz, Fitzgerald, Putt, and
  Ubel}{Armstrong et~al\mbox{.}}{2002}]%
        {armstrong2002effect}
\bibfield{author}{\bibinfo{person}{Katrina Armstrong},
  \bibinfo{person}{J~Sanford Schwartz}, \bibinfo{person}{Genevieve Fitzgerald},
  \bibinfo{person}{Mary Putt}, {and} \bibinfo{person}{Peter~A Ubel}.}
  \bibinfo{year}{2002}\natexlab{}.
\newblock \showarticletitle{Effect of framing as gain versus loss on
  understanding and hypothetical treatment choices: survival and mortality
  curves}.
\newblock \bibinfo{journal}{\emph{Medical Decision Making}}
  \bibinfo{volume}{22}, \bibinfo{number}{1} (\bibinfo{year}{2002}),
  \bibinfo{pages}{76--83}.
\newblock


\bibitem[\protect\citeauthoryear{Atassi, Berry, Shui, Zach, Sherman, Sinani,
  Walker, Katsovskiy, Schoenfeld, and Cudkowicz}{Atassi et~al\mbox{.}}{2014}]%
        {atassiPROACTDatabaseDesign2014}
\bibfield{author}{\bibinfo{person}{Nazem Atassi}, \bibinfo{person}{James
  Berry}, \bibinfo{person}{Amy Shui}, \bibinfo{person}{Neta Zach},
  \bibinfo{person}{Alexander Sherman}, \bibinfo{person}{Ervin Sinani},
  \bibinfo{person}{Jason Walker}, \bibinfo{person}{Igor Katsovskiy},
  \bibinfo{person}{David Schoenfeld}, {and} \bibinfo{person}{Merit Cudkowicz}.}
  \bibinfo{year}{2014}\natexlab{}.
\newblock \showarticletitle{The {{PRO}}-{{ACT}} Database {{Design}}, Initial
  Analyses, and Predictive Features}.
\newblock \bibinfo{journal}{\emph{Neurology}} \bibinfo{volume}{83},
  \bibinfo{number}{19} (\bibinfo{year}{2014}), \bibinfo{pages}{1719--1725}.
\newblock


\bibitem[\protect\citeauthoryear{Atkins, Droegemeier, Feldman, Garcia-Molina,
  Klein, Messerschmitt, Messina, Ostriker, and Wright}{Atkins
  et~al\mbox{.}}{2003}]%
        {atkins2003revolutionizing}
\bibfield{author}{\bibinfo{person}{Daniel~E Atkins}, \bibinfo{person}{Kelvin~K
  Droegemeier}, \bibinfo{person}{Stuart~I Feldman}, \bibinfo{person}{Hector
  Garcia-Molina}, \bibinfo{person}{Michael~L Klein}, \bibinfo{person}{David~G
  Messerschmitt}, \bibinfo{person}{Paul Messina}, \bibinfo{person}{Jeremiah~P
  Ostriker}, {and} \bibinfo{person}{Margaret~H Wright}.}
  \bibinfo{year}{2003}\natexlab{}.
\newblock \showarticletitle{Revolutionizing science and engineering through
  cyberinfrastructure}.
\newblock \bibinfo{journal}{\emph{Report of the National Science Foundation
  blue-ribbon advisory panel on cyberinfrastructure}}  \bibinfo{volume}{1}
  (\bibinfo{year}{2003}).
\newblock


\bibitem[\protect\citeauthoryear{Atkinson and Flint}{Atkinson and
  Flint}{2001}]%
        {atkinson2001accessing}
\bibfield{author}{\bibinfo{person}{Rowland Atkinson} {and}
  \bibinfo{person}{John Flint}.} \bibinfo{year}{2001}\natexlab{}.
\newblock \showarticletitle{Accessing hidden and hard-to-reach populations:
  Snowball research strategies}.
\newblock \bibinfo{journal}{\emph{Social research update}}
  \bibinfo{volume}{33}, \bibinfo{number}{1} (\bibinfo{year}{2001}),
  \bibinfo{pages}{1--4}.
\newblock


\bibitem[\protect\citeauthoryear{Beers, Boshuizen, Kirschner, and
  Gijselaers}{Beers et~al\mbox{.}}{2006}]%
        {beers2006common}
\bibfield{author}{\bibinfo{person}{Pieter~J Beers}, \bibinfo{person}{Henny~PA
  Boshuizen}, \bibinfo{person}{Paul~A Kirschner}, {and} \bibinfo{person}{Wim~H
  Gijselaers}.} \bibinfo{year}{2006}\natexlab{}.
\newblock \showarticletitle{Common ground, complex problems and decision
  making}.
\newblock \bibinfo{journal}{\emph{Group Decision and Negotiation}}
  \bibinfo{volume}{15}, \bibinfo{number}{6} (\bibinfo{year}{2006}),
  \bibinfo{pages}{529--556}.
\newblock


\bibitem[\protect\citeauthoryear{Bhabha}{Bhabha}{1994}]%
        {bhabha1994}
\bibfield{author}{\bibinfo{person}{H Bhabha}.} \bibinfo{year}{1994}\natexlab{}.
\newblock \showarticletitle{The location of culture. London: Routledge.}
\newblock  (\bibinfo{year}{1994}).
\newblock


\bibitem[\protect\citeauthoryear{Bietz, Abrams, Cooper, Stevens, Puga, Patel,
  Olson, and Olson}{Bietz et~al\mbox{.}}{2012}]%
        {bietz2012improving}
\bibfield{author}{\bibinfo{person}{Matthew~J Bietz}, \bibinfo{person}{Steve
  Abrams}, \bibinfo{person}{Dan~M Cooper}, \bibinfo{person}{Kathleen~R
  Stevens}, \bibinfo{person}{Frank Puga}, \bibinfo{person}{Darpan~I Patel},
  \bibinfo{person}{Gary~M Olson}, {and} \bibinfo{person}{Judith~S Olson}.}
  \bibinfo{year}{2012}\natexlab{}.
\newblock \showarticletitle{Improving the odds through the Collaboration
  Success Wizard}.
\newblock \bibinfo{journal}{\emph{Translational behavioral medicine}}
  \bibinfo{volume}{2}, \bibinfo{number}{4} (\bibinfo{year}{2012}),
  \bibinfo{pages}{480--486}.
\newblock


\bibitem[\protect\citeauthoryear{Birnholtz and Bietz}{Birnholtz and
  Bietz}{2003}]%
        {birnholtz2003data}
\bibfield{author}{\bibinfo{person}{Jeremy~P Birnholtz} {and}
  \bibinfo{person}{Matthew~J Bietz}.} \bibinfo{year}{2003}\natexlab{}.
\newblock \showarticletitle{Data at work: supporting sharing in science and
  engineering}. In \bibinfo{booktitle}{\emph{Proceedings of the 2003
  international ACM SIGGROUP conference on Supporting group work}}. ACM,
  \bibinfo{pages}{339--348}.
\newblock


\bibitem[\protect\citeauthoryear{Birnholtz and Finholt}{Birnholtz and
  Finholt}{2013}]%
        {birnholtz2013cultural}
\bibfield{author}{\bibinfo{person}{Jeremy~P Birnholtz} {and}
  \bibinfo{person}{Thomas~A Finholt}.} \bibinfo{year}{2013}\natexlab{}.
\newblock \showarticletitle{Cultural challenges to leadership in
  cyberinfrastructure development}.
\newblock \bibinfo{journal}{\emph{Leadership at a distance: research in
  technologically-supported work}} (\bibinfo{year}{2013}),
  \bibinfo{pages}{195}.
\newblock


\bibitem[\protect\citeauthoryear{Bj{\o}rn, Esbensen, Jensen, and
  Matthiesen}{Bj{\o}rn et~al\mbox{.}}{2014}]%
        {bjornDoesDistanceStill2014a}
\bibfield{author}{\bibinfo{person}{Pernille Bj{\o}rn}, \bibinfo{person}{Morten
  Esbensen}, \bibinfo{person}{Rasmus~Eskild Jensen}, {and}
  \bibinfo{person}{Stina Matthiesen}.} \bibinfo{year}{2014}\natexlab{}.
\newblock \showarticletitle{Does {{Distance Still Matter}}? {{Revisiting}} the
  {{CSCW Fundamentals}} on {{Distributed Collaboration}}}.
\newblock \bibinfo{journal}{\emph{ACM Transactions on Computer-Human
  Interaction}} \bibinfo{volume}{21}, \bibinfo{number}{5} (\bibinfo{date}{Nov.}
  \bibinfo{year}{2014}), \bibinfo{pages}{1--26}.
\newblock
\showISSN{10730516}
\urldef\tempurl%
\url{https://doi.org/10.1145/2670534}
\showDOI{\tempurl}


\bibitem[\protect\citeauthoryear{B{\o}dker, Ehn, Knudsen, Kyng, and
  Madsen}{B{\o}dker et~al\mbox{.}}{1988}]%
        {bodker1988computer}
\bibfield{author}{\bibinfo{person}{Susanne B{\o}dker}, \bibinfo{person}{Pelle
  Ehn}, \bibinfo{person}{Joergen Knudsen}, \bibinfo{person}{Morten Kyng}, {and}
  \bibinfo{person}{Kim Madsen}.} \bibinfo{year}{1988}\natexlab{}.
\newblock \showarticletitle{Computer support for cooperative design}. In
  \bibinfo{booktitle}{\emph{Proceedings of the 1988 ACM conference on
  Computer-supported cooperative work}}. ACM, \bibinfo{pages}{377--394}.
\newblock


\bibitem[\protect\citeauthoryear{Bos, Zimmerman, Olson, Yew, Yerkie, Dahl, and
  Olson}{Bos et~al\mbox{.}}{2007}]%
        {bosSharedDatabasesCommunities2007}
\bibfield{author}{\bibinfo{person}{Nathan Bos}, \bibinfo{person}{Ann
  Zimmerman}, \bibinfo{person}{Judith Olson}, \bibinfo{person}{Jude Yew},
  \bibinfo{person}{Jason Yerkie}, \bibinfo{person}{Erik Dahl}, {and}
  \bibinfo{person}{Gary Olson}.} \bibinfo{year}{2007}\natexlab{}.
\newblock \showarticletitle{From Shared Databases to Communities of Practice:
  {{A}} Taxonomy of Collaboratories}.
\newblock \bibinfo{journal}{\emph{Journal of Computer-Mediated Communication}}
  \bibinfo{volume}{12}, \bibinfo{number}{2} (\bibinfo{year}{2007}),
  \bibinfo{pages}{652--672}.
\newblock


\bibitem[\protect\citeauthoryear{Boyatzis}{Boyatzis}{1998}]%
        {boyatzis1998transforming}
\bibfield{author}{\bibinfo{person}{Richard~E Boyatzis}.}
  \bibinfo{year}{1998}\natexlab{}.
\newblock \bibinfo{booktitle}{\emph{Transforming qualitative information:
  Thematic analysis and code development}}.
\newblock \bibinfo{publisher}{sage}.
\newblock


\bibitem[\protect\citeauthoryear{Carpenter}{Carpenter}{2011}]%
        {carpenter2011may}
\bibfield{author}{\bibinfo{person}{Jennifer Carpenter}.}
  \bibinfo{year}{2011}\natexlab{}.
\newblock \bibinfo{title}{May the best analyst win}.
\newblock
\newblock


\bibitem[\protect\citeauthoryear{CASP13}{CASP13}{2018}]%
        {CASP13}
\bibfield{author}{\bibinfo{person}{CASP13}.} \bibinfo{year}{2018}\natexlab{}.
\newblock \bibinfo{title}{13th Community Wide Experiment on the Critical
  Assessment of Techniques for Protein Structure Prediction}.
\newblock
\newblock
\urldef\tempurl%
\url{http://predictioncenter.org/casp13/}
\showURL{%
\tempurl}


\bibitem[\protect\citeauthoryear{CERN}{CERN}{2018}]%
        {cern}
\bibfield{author}{\bibinfo{person}{CERN}.} \bibinfo{year}{2018}\natexlab{}.
\newblock \bibinfo{title}{CERN Annual report 2017}.
\newblock
\newblock
\urldef\tempurl%
\url{https://cds.cern.ch/record/2624296/files/18030409_CERN_rapport_2017EN.pdf}
\showURL{%
\tempurl}


\bibitem[\protect\citeauthoryear{Clark}{Clark}{1996}]%
        {clark1996using}
\bibfield{author}{\bibinfo{person}{Herbert~H Clark}.}
  \bibinfo{year}{1996}\natexlab{}.
\newblock \bibinfo{booktitle}{\emph{Using language}}.
\newblock \bibinfo{publisher}{Cambridge university press}.
\newblock


\bibitem[\protect\citeauthoryear{Clark, Brennan, et~al\mbox{.}}{Clark
  et~al\mbox{.}}{1991}]%
        {clark1991grounding}
\bibfield{author}{\bibinfo{person}{Herbert~H Clark}, \bibinfo{person}{Susan~E
  Brennan}, {et~al\mbox{.}}} \bibinfo{year}{1991}\natexlab{}.
\newblock \showarticletitle{Grounding in communication}.
\newblock \bibinfo{journal}{\emph{Perspectives on socially shared cognition}}
  \bibinfo{volume}{13}, \bibinfo{number}{1991} (\bibinfo{year}{1991}),
  \bibinfo{pages}{127--149}.
\newblock


\bibitem[\protect\citeauthoryear{Convertino, Ganoe, Schafer, Yost, and
  Carroll}{Convertino et~al\mbox{.}}{2005}]%
        {convertino2005multiple}
\bibfield{author}{\bibinfo{person}{Gregorio Convertino},
  \bibinfo{person}{Craig~H Ganoe}, \bibinfo{person}{Wendy~A Schafer},
  \bibinfo{person}{Beth Yost}, {and} \bibinfo{person}{John~M Carroll}.}
  \bibinfo{year}{2005}\natexlab{}.
\newblock \showarticletitle{A multiple view approach to support common ground
  in distributed and synchronous geo-collaboration}. In
  \bibinfo{booktitle}{\emph{Coordinated and Multiple Views in Exploratory
  Visualization (CMV'05)}}. IEEE, \bibinfo{pages}{121--132}.
\newblock


\bibitem[\protect\citeauthoryear{Convertino, Mentis, Rosson, Carroll,
  Slavkovic, and Ganoe}{Convertino et~al\mbox{.}}{2008}]%
        {convertino2008articulating}
\bibfield{author}{\bibinfo{person}{Gregorio Convertino},
  \bibinfo{person}{Helena~M Mentis}, \bibinfo{person}{Mary~Beth Rosson},
  \bibinfo{person}{John~M Carroll}, \bibinfo{person}{Aleksandra Slavkovic},
  {and} \bibinfo{person}{Craig~H Ganoe}.} \bibinfo{year}{2008}\natexlab{}.
\newblock \showarticletitle{Articulating common ground in cooperative work:
  content and process}. In \bibinfo{booktitle}{\emph{Proceedings of the SIGCHI
  conference on human factors in computing systems}}. ACM,
  \bibinfo{pages}{1637--1646}.
\newblock


\bibitem[\protect\citeauthoryear{Convertino, Mentis, Ting, Rosson, and
  Carroll}{Convertino et~al\mbox{.}}{2007}]%
        {convertino2007does}
\bibfield{author}{\bibinfo{person}{Gregorio Convertino},
  \bibinfo{person}{Helena~M Mentis}, \bibinfo{person}{Alex~YW Ting},
  \bibinfo{person}{Mary~Beth Rosson}, {and} \bibinfo{person}{John~M Carroll}.}
  \bibinfo{year}{2007}\natexlab{}.
\newblock \showarticletitle{How does common ground increase?}. In
  \bibinfo{booktitle}{\emph{Proceedings of the 2007 international ACM
  conference on Supporting group work}}. ACM, \bibinfo{pages}{225--228}.
\newblock


\bibitem[\protect\citeauthoryear{Crabtree and Miller}{Crabtree and
  Miller}{1999}]%
        {crabtree1999doing}
\bibfield{author}{\bibinfo{person}{Benjamin~F Crabtree} {and}
  \bibinfo{person}{William~L Miller}.} \bibinfo{year}{1999}\natexlab{}.
\newblock \bibinfo{booktitle}{\emph{Doing qualitative research}}.
\newblock \bibinfo{publisher}{sage publications}.
\newblock


\bibitem[\protect\citeauthoryear{Creswell and Creswell}{Creswell and
  Creswell}{2017}]%
        {creswell2017research}
\bibfield{author}{\bibinfo{person}{John~W Creswell} {and}
  \bibinfo{person}{J~David Creswell}.} \bibinfo{year}{2017}\natexlab{}.
\newblock \bibinfo{booktitle}{\emph{Research design: Qualitative, quantitative,
  and mixed methods approaches}}.
\newblock \bibinfo{publisher}{Sage publications}.
\newblock


\bibitem[\protect\citeauthoryear{Dabbish, Stuart, Tsay, and Herbsleb}{Dabbish
  et~al\mbox{.}}{2012}]%
        {dabbishSocialCodingGitHub2012}
\bibfield{author}{\bibinfo{person}{Laura Dabbish}, \bibinfo{person}{Colleen
  Stuart}, \bibinfo{person}{Jason Tsay}, {and} \bibinfo{person}{Jim Herbsleb}.}
  \bibinfo{year}{2012}\natexlab{}.
\newblock \showarticletitle{Social Coding in {{GitHub}}: Transparency and
  Collaboration in an Open Software Repository}. In
  \bibinfo{booktitle}{\emph{Proceedings of the {{ACM}} 2012 Conference on
  {{Computer Supported Cooperative Work}}}}. \bibinfo{publisher}{{ACM}},
  \bibinfo{pages}{1277--1286}.
\newblock


\bibitem[\protect\citeauthoryear{data.gov}{data.gov}{2019}]%
        {datagov}
\bibfield{author}{\bibinfo{person}{data.gov}.} \bibinfo{year}{2019}\natexlab{}.
\newblock \bibinfo{title}{data.gove datasets}.
\newblock
\newblock
\urldef\tempurl%
\url{https://catalog.data.gov/dataset}
\showURL{%
\tempurl}


\bibitem[\protect\citeauthoryear{datarobot}{datarobot}{2019}]%
        {datarobot}
\bibfield{author}{\bibinfo{person}{datarobot}.}
  \bibinfo{year}{2019}\natexlab{}.
\newblock \bibinfo{title}{datarobot}.
\newblock
\newblock
\urldef\tempurl%
\url{https://www.datarobot.com/}
\showURL{%
\tempurl}


\bibitem[\protect\citeauthoryear{Derry, Schunn, and Gernsbacher}{Derry
  et~al\mbox{.}}{2014}]%
        {derry2014interdisciplinary}
\bibfield{author}{\bibinfo{person}{Sharon~J Derry},
  \bibinfo{person}{Christian~D Schunn}, {and} \bibinfo{person}{Morton~Ann
  Gernsbacher}.} \bibinfo{year}{2014}\natexlab{}.
\newblock \bibinfo{booktitle}{\emph{Interdisciplinary collaboration: An
  emerging cognitive science}}.
\newblock \bibinfo{publisher}{Psychology Press}.
\newblock


\bibitem[\protect\citeauthoryear{Dourish and Bellotti}{Dourish and
  Bellotti}{1992}]%
        {dourish1992awareness}
\bibfield{author}{\bibinfo{person}{Paul Dourish} {and}
  \bibinfo{person}{Victoria Bellotti}.} \bibinfo{year}{1992}\natexlab{}.
\newblock \showarticletitle{Awareness and coordination in shared workspaces.}.
  In \bibinfo{booktitle}{\emph{CSCW}}, Vol.~\bibinfo{volume}{92}.
  \bibinfo{pages}{107--114}.
\newblock


\bibitem[\protect\citeauthoryear{Fiedler, Juslin, et~al\mbox{.}}{Fiedler
  et~al\mbox{.}}{2006}]%
        {fiedler2006information}
\bibfield{author}{\bibinfo{person}{Klaus Fiedler}, \bibinfo{person}{Peter
  Juslin}, {et~al\mbox{.}}} \bibinfo{year}{2006}\natexlab{}.
\newblock \bibinfo{booktitle}{\emph{Information sampling and adaptive
  cognition}}.
\newblock \bibinfo{publisher}{Cambridge University Press}.
\newblock


\bibitem[\protect\citeauthoryear{Filla}{Filla}{2018}]%
        {watson}
\bibfield{author}{\bibinfo{person}{Greg Filla}.}
  \bibinfo{year}{2018}\natexlab{}.
\newblock \bibinfo{title}{What's New with Watson Machine Learning?}
\newblock
\newblock
\urldef\tempurl%
\url{https://medium.com/ibm-watson/whats-new-with-watson-machine-learning-4de86aa1469d}
\showURL{%
\tempurl}


\bibitem[\protect\citeauthoryear{Gamliel and Peer}{Gamliel and Peer}{2010}]%
        {gamliel2010attribute}
\bibfield{author}{\bibinfo{person}{Eyal Gamliel} {and} \bibinfo{person}{Eyal
  Peer}.} \bibinfo{year}{2010}\natexlab{}.
\newblock \showarticletitle{Attribute framing affects the perceived fairness of
  health care allocation principles}.
\newblock \bibinfo{journal}{\emph{Judgment and Decision Making}}
  \bibinfo{volume}{5}, \bibinfo{number}{1} (\bibinfo{year}{2010}),
  \bibinfo{pages}{11}.
\newblock


\bibitem[\protect\citeauthoryear{GenBank}{GenBank}{2019}]%
        {genbank}
\bibfield{author}{\bibinfo{person}{GenBank}.} \bibinfo{year}{2019}\natexlab{}.
\newblock \bibinfo{title}{GenBank Statistics}.
\newblock
\newblock
\urldef\tempurl%
\url{https://www.ncbi.nlm.nih.gov/genbank/statistics/}
\showURL{%
\tempurl}


\bibitem[\protect\citeauthoryear{Gil, Greaves, Hendler, and Hirsh}{Gil
  et~al\mbox{.}}{2014}]%
        {gil2014amplify}
\bibfield{author}{\bibinfo{person}{Yolanda Gil}, \bibinfo{person}{Mark
  Greaves}, \bibinfo{person}{James Hendler}, {and} \bibinfo{person}{Haym
  Hirsh}.} \bibinfo{year}{2014}\natexlab{}.
\newblock \showarticletitle{Amplify scientific discovery with artificial
  intelligence}.
\newblock \bibinfo{journal}{\emph{Science}} \bibinfo{volume}{346},
  \bibinfo{number}{6206} (\bibinfo{year}{2014}), \bibinfo{pages}{171--172}.
\newblock


\bibitem[\protect\citeauthoryear{Google}{Google}{2019}]%
        {google}
\bibfield{author}{\bibinfo{person}{Google}.} \bibinfo{year}{2019}\natexlab{}.
\newblock \bibinfo{title}{Cloud AutoML}.
\newblock
\newblock
\urldef\tempurl%
\url{https://cloud.google.com/automl/}
\showURL{%
\tempurl}


\bibitem[\protect\citeauthoryear{Gorman}{Gorman}{2002}]%
        {gorman2002expanding}
\bibfield{author}{\bibinfo{person}{Michael~E Gorman}.}
  \bibinfo{year}{2002}\natexlab{}.
\newblock \showarticletitle{Expanding the trading zones for convergent
  technologies}.
\newblock \bibinfo{journal}{\emph{Converging Technologies for Improving Human
  Performance}} (\bibinfo{year}{2002}), \bibinfo{pages}{424}.
\newblock


\bibitem[\protect\citeauthoryear{Gorman}{Gorman}{2008}]%
        {gorman2008scientific}
\bibfield{author}{\bibinfo{person}{Michael~E Gorman}.}
  \bibinfo{year}{2008}\natexlab{}.
\newblock \showarticletitle{Scientific and technological expertise.}
\newblock \bibinfo{journal}{\emph{Journal of psychology of science and
  technology}} (\bibinfo{year}{2008}).
\newblock


\bibitem[\protect\citeauthoryear{Gray}{Gray}{1989}]%
        {gray1989collaborating}
\bibfield{author}{\bibinfo{person}{Barbara Gray}.}
  \bibinfo{year}{1989}\natexlab{}.
\newblock \showarticletitle{Collaborating: Finding common ground for multiparty
  problems}.
\newblock  (\bibinfo{year}{1989}).
\newblock


\bibitem[\protect\citeauthoryear{Greenberg}{Greenberg}{1990}]%
        {greenberg1990sharing}
\bibfield{author}{\bibinfo{person}{Saul Greenberg}.}
  \bibinfo{year}{1990}\natexlab{}.
\newblock \showarticletitle{Sharing views and interactions with single-user
  applications}. In \bibinfo{booktitle}{\emph{ACM SIGOIS Bulletin}},
  Vol.~\bibinfo{volume}{11}. ACM, \bibinfo{pages}{227--237}.
\newblock


\bibitem[\protect\citeauthoryear{H2O}{H2O}{2019}]%
        {H2O}
\bibfield{author}{\bibinfo{person}{H2O}.} \bibinfo{year}{2019}\natexlab{}.
\newblock \bibinfo{title}{H2O.ai}.
\newblock
\newblock
\urldef\tempurl%
\url{https://www.h2o.ai/}
\showURL{%
\tempurl}


\bibitem[\protect\citeauthoryear{Hill, Bellamy, Erickson, and Burnett}{Hill
  et~al\mbox{.}}{2016}]%
        {hillTrialsTribulationsDevelopers2016}
\bibfield{author}{\bibinfo{person}{Charles Hill}, \bibinfo{person}{Rachel
  Bellamy}, \bibinfo{person}{Thomas Erickson}, {and} \bibinfo{person}{Margaret
  Burnett}.} \bibinfo{year}{2016}\natexlab{}.
\newblock \showarticletitle{Trials and Tribulations of Developers of
  Intelligent Systems: {{A}} Field Study}. In \bibinfo{booktitle}{\emph{Visual
  {{Languages}} and {{Human}}-{{Centric Computing}} ({{VL}}/{{HCC}}), 2016
  {{IEEE Symposium}} On}}. \bibinfo{publisher}{{IEEE}},
  \bibinfo{pages}{162--170}.
\newblock


\bibitem[\protect\citeauthoryear{Hou and Wang}{Hou and Wang}{2017}]%
        {hou2017hacking}
\bibfield{author}{\bibinfo{person}{Youyang Hou} {and} \bibinfo{person}{Dakuo
  Wang}.} \bibinfo{year}{2017}\natexlab{}.
\newblock \showarticletitle{Hacking with NPOs: collaborative analytics and
  broker roles in civic data hackathons}.
\newblock \bibinfo{journal}{\emph{Proceedings of the ACM on Human-Computer
  Interaction}} \bibinfo{volume}{1}, \bibinfo{number}{CSCW}
  (\bibinfo{year}{2017}), \bibinfo{pages}{53}.
\newblock


\bibitem[\protect\citeauthoryear{Hutchins}{Hutchins}{1995}]%
        {hutchins1995cockpit}
\bibfield{author}{\bibinfo{person}{Edwin Hutchins}.}
  \bibinfo{year}{1995}\natexlab{}.
\newblock \showarticletitle{How a cockpit remembers its speeds}.
\newblock \bibinfo{journal}{\emph{Cognitive science}} \bibinfo{volume}{19},
  \bibinfo{number}{3} (\bibinfo{year}{1995}), \bibinfo{pages}{265--288}.
\newblock


\bibitem[\protect\citeauthoryear{Jirotka, Lee, and Olson}{Jirotka
  et~al\mbox{.}}{2013}]%
        {jirotka2013supporting}
\bibfield{author}{\bibinfo{person}{Marina Jirotka},
  \bibinfo{person}{Charlotte~P Lee}, {and} \bibinfo{person}{Gary~M Olson}.}
  \bibinfo{year}{2013}\natexlab{}.
\newblock \showarticletitle{Supporting scientific collaboration: Methods, tools
  and concepts}.
\newblock \bibinfo{journal}{\emph{Computer Supported Cooperative Work (CSCW)}}
  \bibinfo{volume}{22}, \bibinfo{number}{4-6} (\bibinfo{year}{2013}),
  \bibinfo{pages}{667--715}.
\newblock


\bibitem[\protect\citeauthoryear{Kahneman}{Kahneman}{2011}]%
        {kahneman2011thinking}
\bibfield{author}{\bibinfo{person}{Daniel Kahneman}.}
  \bibinfo{year}{2011}\natexlab{}.
\newblock \bibinfo{title}{Thinking fast and slow. Allen Lane}.
\newblock
\newblock


\bibitem[\protect\citeauthoryear{Kandel, Paepcke, Hellerstein, and Heer}{Kandel
  et~al\mbox{.}}{2011}]%
        {kandel2011wrangler}
\bibfield{author}{\bibinfo{person}{Sean Kandel}, \bibinfo{person}{Andreas
  Paepcke}, \bibinfo{person}{Joseph Hellerstein}, {and}
  \bibinfo{person}{Jeffrey Heer}.} \bibinfo{year}{2011}\natexlab{}.
\newblock \showarticletitle{Wrangler: Interactive visual specification of data
  transformation scripts}. In \bibinfo{booktitle}{\emph{Proceedings of the
  SIGCHI Conference on Human Factors in Computing Systems}}. ACM,
  \bibinfo{pages}{3363--3372}.
\newblock


\bibitem[\protect\citeauthoryear{Kandogan, Balakrishnan, Haber, and
  Pierce}{Kandogan et~al\mbox{.}}{2014}]%
        {kandoganDataInsightWork2014}
\bibfield{author}{\bibinfo{person}{Eser Kandogan}, \bibinfo{person}{Aruna
  Balakrishnan}, \bibinfo{person}{Eben~M. Haber}, {and}
  \bibinfo{person}{Jeffrey~S. Pierce}.} \bibinfo{year}{2014}\natexlab{}.
\newblock \showarticletitle{From {{Data}} to {{Insight}}: {{Work Practices}} of
  {{Analysts}} in the {{Enterprise}}}.
\newblock \bibinfo{journal}{\emph{IEEE Computer Graphics and Applications}}
  \bibinfo{volume}{34}, \bibinfo{number}{5} (\bibinfo{date}{Sept.}
  \bibinfo{year}{2014}), \bibinfo{pages}{42--50}.
\newblock
\showISSN{0272-1716}
\urldef\tempurl%
\url{https://doi.org/10.1109/MCG.2014.62}
\showDOI{\tempurl}


\bibitem[\protect\citeauthoryear{Kerr and Tindale}{Kerr and Tindale}{2004}]%
        {kerr2004group}
\bibfield{author}{\bibinfo{person}{Norbert~L Kerr} {and}
  \bibinfo{person}{R~Scott Tindale}.} \bibinfo{year}{2004}\natexlab{}.
\newblock \showarticletitle{Group performance and decision making}.
\newblock \bibinfo{journal}{\emph{Annu. Rev. Psychol.}}  \bibinfo{volume}{55}
  (\bibinfo{year}{2004}), \bibinfo{pages}{623--655}.
\newblock


\bibitem[\protect\citeauthoryear{Khurana, Turaga, Samulowitz, and
  Parthasrathy}{Khurana et~al\mbox{.}}{2016}]%
        {khurana2016cognito}
\bibfield{author}{\bibinfo{person}{Udayan Khurana}, \bibinfo{person}{Deepak
  Turaga}, \bibinfo{person}{Horst Samulowitz}, {and}
  \bibinfo{person}{Srinivasan Parthasrathy}.} \bibinfo{year}{2016}\natexlab{}.
\newblock \showarticletitle{Cognito: Automated feature engineering for
  supervised learning}. In \bibinfo{booktitle}{\emph{2016 IEEE 16th
  International Conference on Data Mining Workshops (ICDMW)}}. IEEE,
  \bibinfo{pages}{1304--1307}.
\newblock


\bibitem[\protect\citeauthoryear{Kirkman, Gibson, and Kim}{Kirkman
  et~al\mbox{.}}{2012}]%
        {kirkman2012across}
\bibfield{author}{\bibinfo{person}{Bradley~L Kirkman},
  \bibinfo{person}{Cristina~B Gibson}, {and} \bibinfo{person}{Kwanghyun Kim}.}
  \bibinfo{year}{2012}\natexlab{}.
\newblock \showarticletitle{Across borders and technologies: Advancements in
  virtual teams research}.
\newblock In \bibinfo{booktitle}{\emph{The Oxford Handbook of Organizational
  Psychology, Volume 2}}.
\newblock


\bibitem[\protect\citeauthoryear{Kluyver, Ragan-Kelley, P{\'e}rez, Granger,
  Bussonnier, Frederic, Kelley, Hamrick, Grout, Corlay, et~al\mbox{.}}{Kluyver
  et~al\mbox{.}}{2016}]%
        {kluyver2016jupyter}
\bibfield{author}{\bibinfo{person}{Thomas Kluyver}, \bibinfo{person}{Benjamin
  Ragan-Kelley}, \bibinfo{person}{Fernando P{\'e}rez}, \bibinfo{person}{Brian~E
  Granger}, \bibinfo{person}{Matthias Bussonnier}, \bibinfo{person}{Jonathan
  Frederic}, \bibinfo{person}{Kyle Kelley}, \bibinfo{person}{Jessica~B
  Hamrick}, \bibinfo{person}{Jason Grout}, \bibinfo{person}{Sylvain Corlay},
  {et~al\mbox{.}}} \bibinfo{year}{2016}\natexlab{}.
\newblock \showarticletitle{Jupyter Notebooks-a publishing format for
  reproducible computational workflows.}. In \bibinfo{booktitle}{\emph{ELPUB}}.
  \bibinfo{pages}{87--90}.
\newblock


\bibitem[\protect\citeauthoryear{Kruglanski and Webster}{Kruglanski and
  Webster}{1996}]%
        {kruglanski1996motivated}
\bibfield{author}{\bibinfo{person}{Arie~W Kruglanski} {and} \bibinfo{person}{DM
  Webster}.} \bibinfo{year}{1996}\natexlab{}.
\newblock \showarticletitle{Motivated closing of the mind: Its cognitive and
  social effects}.
\newblock \bibinfo{journal}{\emph{Psychological Review}} \bibinfo{volume}{103},
  \bibinfo{number}{2} (\bibinfo{year}{1996}), \bibinfo{pages}{263--283}.
\newblock


\bibitem[\protect\citeauthoryear{K{\"u}ffner, Zach, Norel, Hawe, Schoenfeld,
  Wang, Li, Fang, Mackey, Hardiman, et~al\mbox{.}}{K{\"u}ffner
  et~al\mbox{.}}{2015}]%
        {kuffner2015crowdsourced}
\bibfield{author}{\bibinfo{person}{Robert K{\"u}ffner}, \bibinfo{person}{Neta
  Zach}, \bibinfo{person}{Raquel Norel}, \bibinfo{person}{Johann Hawe},
  \bibinfo{person}{David Schoenfeld}, \bibinfo{person}{Liuxia Wang},
  \bibinfo{person}{Guang Li}, \bibinfo{person}{Lilly Fang},
  \bibinfo{person}{Lester Mackey}, \bibinfo{person}{Orla Hardiman},
  {et~al\mbox{.}}} \bibinfo{year}{2015}\natexlab{}.
\newblock \showarticletitle{Crowdsourced analysis of clinical trial data to
  predict amyotrophic lateral sclerosis progression}.
\newblock \bibinfo{journal}{\emph{Nature biotechnology}} \bibinfo{volume}{33},
  \bibinfo{number}{1} (\bibinfo{year}{2015}), \bibinfo{pages}{51}.
\newblock


\bibitem[\protect\citeauthoryear{K{\"u}hberger}{K{\"u}hberger}{1998}]%
        {kuhberger1998influence}
\bibfield{author}{\bibinfo{person}{Anton K{\"u}hberger}.}
  \bibinfo{year}{1998}\natexlab{}.
\newblock \showarticletitle{The influence of framing on risky decisions: A
  meta-analysis}.
\newblock \bibinfo{journal}{\emph{Organizational behavior and human decision
  processes}} \bibinfo{volume}{75}, \bibinfo{number}{1} (\bibinfo{year}{1998}),
  \bibinfo{pages}{23--55}.
\newblock


\bibitem[\protect\citeauthoryear{Lawrence}{Lawrence}{2006}]%
        {lawrenceWalkingTightropeBalancing2006}
\bibfield{author}{\bibinfo{person}{Katherine~A. Lawrence}.}
  \bibinfo{year}{2006}\natexlab{}.
\newblock \showarticletitle{Walking the {{Tightrope}}: {{The Balancing Acts}}
  of a {{Large}} e-{{Research Project}}}.
\newblock \bibinfo{journal}{\emph{Computer Supported Cooperative Work (CSCW)}}
  \bibinfo{volume}{15}, \bibinfo{number}{4} (\bibinfo{date}{Oct.}
  \bibinfo{year}{2006}), \bibinfo{pages}{385--411}.
\newblock
\showISSN{0925-9724, 1573-7551}
\urldef\tempurl%
\url{https://doi.org/10.1007/s10606-006-9025-0}
\showDOI{\tempurl}


\bibitem[\protect\citeauthoryear{Lee, Bietz, and Thayer}{Lee
  et~al\mbox{.}}{2010}]%
        {lee2010driven}
\bibfield{author}{\bibinfo{person}{Charlotte~P Lee}, \bibinfo{person}{Matthew~J
  Bietz}, {and} \bibinfo{person}{Alexander Thayer}.}
  \bibinfo{year}{2010}\natexlab{}.
\newblock \showarticletitle{Research-driven stakeholders in cyberinfrastructure
  use and development}. In \bibinfo{booktitle}{\emph{2010 International
  Symposium on Collaborative Technologies and Systems}}. IEEE,
  \bibinfo{pages}{163--172}.
\newblock


\bibitem[\protect\citeauthoryear{Liu, Ram, Bouneffouf, Vijaykeerthy, Bramble,
  Samulowitz, Wang, Conn, and Gray}{Liu et~al\mbox{.}}{2019}]%
        {liu2019formal}
\bibfield{author}{\bibinfo{person}{Sijia Liu}, \bibinfo{person}{Parikshit Ram},
  \bibinfo{person}{Djallel Bouneffouf}, \bibinfo{person}{Deepak Vijaykeerthy},
  \bibinfo{person}{Gregory Bramble}, \bibinfo{person}{Horst Samulowitz},
  \bibinfo{person}{Dakuo Wang}, \bibinfo{person}{Andrew~R Conn}, {and}
  \bibinfo{person}{Alexander Gray}.} \bibinfo{year}{2019}\natexlab{}.
\newblock \bibinfo{title}{A Formal Method for AutoML via ADMM}.
\newblock
\newblock
\showeprint[arxiv]{1905.00424}


\bibitem[\protect\citeauthoryear{Luo, Ng'ambi, and Hanss}{Luo
  et~al\mbox{.}}{2010}]%
        {luo2010towards}
\bibfield{author}{\bibinfo{person}{Airong Luo}, \bibinfo{person}{Dick Ng'ambi},
  {and} \bibinfo{person}{Ted Hanss}.} \bibinfo{year}{2010}\natexlab{}.
\newblock \showarticletitle{Towards building a productive, scalable and
  sustainable collaboration model for open educational resources}. In
  \bibinfo{booktitle}{\emph{Proceedings of the 16th ACM international
  conference on Supporting group work}}. ACM, \bibinfo{pages}{273--282}.
\newblock


\bibitem[\protect\citeauthoryear{Mesmer-Magnus and DeChurch}{Mesmer-Magnus and
  DeChurch}{2009}]%
        {mesmer2009information}
\bibfield{author}{\bibinfo{person}{Jessica~R Mesmer-Magnus} {and}
  \bibinfo{person}{Leslie~A DeChurch}.} \bibinfo{year}{2009}\natexlab{}.
\newblock \showarticletitle{Information sharing and team performance: A
  meta-analysis.}
\newblock \bibinfo{journal}{\emph{Journal of Applied Psychology}}
  \bibinfo{volume}{94}, \bibinfo{number}{2} (\bibinfo{year}{2009}),
  \bibinfo{pages}{535}.
\newblock


\bibitem[\protect\citeauthoryear{Monk}{Monk}{2003}]%
        {monk2003common}
\bibfield{author}{\bibinfo{person}{Andrew Monk}.}
  \bibinfo{year}{2003}\natexlab{}.
\newblock \showarticletitle{Common ground in electronically mediated
  communication: Clark's theory of language use}.
\newblock \bibinfo{journal}{\emph{HCI models, theories, and frameworks: Toward
  a multidisciplinary science}} (\bibinfo{year}{2003}),
  \bibinfo{pages}{265--289}.
\newblock


\bibitem[\protect\citeauthoryear{Muller, Lange, Wang, Piorkowski, Tsay, Liao,
  Dugan, and Erickson}{Muller et~al\mbox{.}}{2019}]%
        {muller2019data}
\bibfield{author}{\bibinfo{person}{Michael Muller}, \bibinfo{person}{Ingrid
  Lange}, \bibinfo{person}{Dakuo Wang}, \bibinfo{person}{David Piorkowski},
  \bibinfo{person}{Jason Tsay}, \bibinfo{person}{Q~Vera Liao},
  \bibinfo{person}{Casey Dugan}, {and} \bibinfo{person}{Thomas Erickson}.}
  \bibinfo{year}{2019}\natexlab{}.
\newblock \showarticletitle{How Data Science Workers Work with Data: Discovery,
  Capture, Curation, Design, Creation}. In
  \bibinfo{booktitle}{\emph{Proceedings of the 2019 CHI Conference on Human
  Factors in Computing Systems}}. ACM, \bibinfo{pages}{126}.
\newblock


\bibitem[\protect\citeauthoryear{Muller and Druin}{Muller and Druin}{2010}]%
        {muller2010participatory}
\bibfield{author}{\bibinfo{person}{Michael~J Muller} {and}
  \bibinfo{person}{Allison Druin}.} \bibinfo{year}{2010}\natexlab{}.
\newblock \showarticletitle{Participatory design: the third space in hci.
  human-computer interaction: Development process. J. Jacko and A}.
\newblock \bibinfo{journal}{\emph{Sears. Eds. Handbook of HCI}}
  (\bibinfo{year}{2010}).
\newblock


\bibitem[\protect\citeauthoryear{Nargesian, Samulowitz, Khurana, Khalil, and
  Turaga}{Nargesian et~al\mbox{.}}{2017}]%
        {nargesian2017learning}
\bibfield{author}{\bibinfo{person}{Fatemeh Nargesian}, \bibinfo{person}{Horst
  Samulowitz}, \bibinfo{person}{Udayan Khurana}, \bibinfo{person}{Elias~B
  Khalil}, {and} \bibinfo{person}{Deepak~S Turaga}.}
  \bibinfo{year}{2017}\natexlab{}.
\newblock \showarticletitle{Learning Feature Engineering for Classification.}.
  In \bibinfo{booktitle}{\emph{IJCAI}}. \bibinfo{pages}{2529--2535}.
\newblock


\bibitem[\protect\citeauthoryear{Oldenburg and Brissett}{Oldenburg and
  Brissett}{1982}]%
        {oldenburg1982third}
\bibfield{author}{\bibinfo{person}{Ramon Oldenburg} {and}
  \bibinfo{person}{Dennis Brissett}.} \bibinfo{year}{1982}\natexlab{}.
\newblock \showarticletitle{The third place}.
\newblock \bibinfo{journal}{\emph{Qualitative sociology}} \bibinfo{volume}{5},
  \bibinfo{number}{4} (\bibinfo{year}{1982}), \bibinfo{pages}{265--284}.
\newblock


\bibitem[\protect\citeauthoryear{Olson and Olson}{Olson and Olson}{2016}]%
        {olson2016converging}
\bibfield{author}{\bibinfo{person}{GARY~M Olson} {and} \bibinfo{person}{J
  Olson}.} \bibinfo{year}{2016}\natexlab{}.
\newblock \showarticletitle{Converging on theory from four sides}.
\newblock \bibinfo{journal}{\emph{Theory development in the Information
  Sciences. Ed. D. Sonnenwald. Univ. Of Texas, Austin}} (\bibinfo{year}{2016}),
  \bibinfo{pages}{87--100}.
\newblock


\bibitem[\protect\citeauthoryear{Olson and Olson}{Olson and Olson}{2000}]%
        {olson2000distance}
\bibfield{author}{\bibinfo{person}{Gary~M Olson} {and}
  \bibinfo{person}{Judith~S Olson}.} \bibinfo{year}{2000}\natexlab{}.
\newblock \showarticletitle{Distance matters}.
\newblock \bibinfo{journal}{\emph{Human--computer interaction}}
  \bibinfo{volume}{15}, \bibinfo{number}{2-3} (\bibinfo{year}{2000}),
  \bibinfo{pages}{139--178}.
\newblock


\bibitem[\protect\citeauthoryear{Olson, Teasley, Bietz, and Cogburn}{Olson
  et~al\mbox{.}}{2002}]%
        {olson2002collaboratories}
\bibfield{author}{\bibinfo{person}{Gary~M Olson}, \bibinfo{person}{Stephanie
  Teasley}, \bibinfo{person}{Matthew~J Bietz}, {and} \bibinfo{person}{Derrick~L
  Cogburn}.} \bibinfo{year}{2002}\natexlab{}.
\newblock \showarticletitle{Collaboratories to support distributed science: the
  example of international HIV/AIDS research}. In
  \bibinfo{booktitle}{\emph{Proceedings of the 2002 annual research conference
  of the South African institute of computer scientists and information
  technologists on enablement through technology}}. South African Institute for
  Computer Scientists and Information Technologists, \bibinfo{pages}{44--51}.
\newblock


\bibitem[\protect\citeauthoryear{Olson, Zimmerman, and Bos}{Olson
  et~al\mbox{.}}{2008}]%
        {olson2008scientific}
\bibfield{author}{\bibinfo{person}{Gary~M Olson}, \bibinfo{person}{Ann
  Zimmerman}, {and} \bibinfo{person}{Nathan Bos}.}
  \bibinfo{year}{2008}\natexlab{}.
\newblock \bibinfo{booktitle}{\emph{Scientific collaboration on the Internet}}.
\newblock \bibinfo{publisher}{The MIT Press}.
\newblock


\bibitem[\protect\citeauthoryear{Olson and Olson}{Olson and Olson}{2013}]%
        {olson2013working}
\bibfield{author}{\bibinfo{person}{Judith~S Olson} {and}
  \bibinfo{person}{Gary~M Olson}.} \bibinfo{year}{2013}\natexlab{}.
\newblock \showarticletitle{Working together apart: Collaboration over the
  internet}.
\newblock \bibinfo{journal}{\emph{Synthesis Lectures on Human-Centered
  Informatics}} \bibinfo{volume}{6}, \bibinfo{number}{5}
  (\bibinfo{year}{2013}), \bibinfo{pages}{1--151}.
\newblock


\bibitem[\protect\citeauthoryear{Olson, Wang, Olson, and Zhang}{Olson
  et~al\mbox{.}}{2017}]%
        {olson2017people}
\bibfield{author}{\bibinfo{person}{Judith~S Olson}, \bibinfo{person}{Dakuo
  Wang}, \bibinfo{person}{Gary~M Olson}, {and} \bibinfo{person}{Jingwen
  Zhang}.} \bibinfo{year}{2017}\natexlab{}.
\newblock \showarticletitle{How people write together now: Beginning the
  investigation with advanced undergraduates in a project course}.
\newblock \bibinfo{journal}{\emph{ACM Transactions on Computer-Human
  Interaction (TOCHI)}} \bibinfo{volume}{24}, \bibinfo{number}{1}
  (\bibinfo{year}{2017}), \bibinfo{pages}{4}.
\newblock


\bibitem[\protect\citeauthoryear{Paepcke}{Paepcke}{1996}]%
        {paepcke1996information}
\bibfield{author}{\bibinfo{person}{Andreas Paepcke}.}
  \bibinfo{year}{1996}\natexlab{}.
\newblock \showarticletitle{Information needs in technical work settings and
  their implications for the design of computer tools}.
\newblock \bibinfo{journal}{\emph{Computer Supported Cooperative Work (CSCW)}}
  \bibinfo{volume}{5}, \bibinfo{number}{1} (\bibinfo{year}{1996}),
  \bibinfo{pages}{63--92}.
\newblock


\bibitem[\protect\citeauthoryear{Paletz and Schunn}{Paletz and Schunn}{2010}]%
        {paletz2010social}
\bibfield{author}{\bibinfo{person}{Susannah~BF Paletz} {and}
  \bibinfo{person}{Christian~D Schunn}.} \bibinfo{year}{2010}\natexlab{}.
\newblock \showarticletitle{A social-cognitive framework of multidisciplinary
  team innovation}.
\newblock \bibinfo{journal}{\emph{Topics in Cognitive Science}}
  \bibinfo{volume}{2}, \bibinfo{number}{1} (\bibinfo{year}{2010}),
  \bibinfo{pages}{73--95}.
\newblock


\bibitem[\protect\citeauthoryear{Patel, Fogarty, Landay, and Harrison}{Patel
  et~al\mbox{.}}{2008}]%
        {patelExaminingDifficultiesSoftware2008}
\bibfield{author}{\bibinfo{person}{Kayur Patel}, \bibinfo{person}{James
  Fogarty}, \bibinfo{person}{James~A. Landay}, {and}
  \bibinfo{person}{Beverly~L. Harrison}.} \bibinfo{year}{2008}\natexlab{}.
\newblock \showarticletitle{Examining {{Difficulties Software Developers
  Encounter}} in the {{Adoption}} of {{Statistical Machine Learning}}.}. In
  \bibinfo{booktitle}{\emph{{{AAAI}}}}. \bibinfo{pages}{1563--1566}.
\newblock


\bibitem[\protect\citeauthoryear{Patterson, Baldini, Mojsilovic, and
  Varshney}{Patterson et~al\mbox{.}}{2018}]%
        {patterson2018semantic}
\bibfield{author}{\bibinfo{person}{Evan Patterson}, \bibinfo{person}{Ioana
  Baldini}, \bibinfo{person}{Aleksandra Mojsilovic}, {and}
  \bibinfo{person}{Kush~R Varshney}.} \bibinfo{year}{2018}\natexlab{}.
\newblock \showarticletitle{Semantic Representation of Data Science Programs.}.
  In \bibinfo{booktitle}{\emph{IJCAI}}. \bibinfo{pages}{5847--5849}.
\newblock


\bibitem[\protect\citeauthoryear{Patterson, McBurney, Schmidt, Baldini,
  Mojsilovi{\'c}, and Varshney}{Patterson et~al\mbox{.}}{2017}]%
        {patterson2017dataflow}
\bibfield{author}{\bibinfo{person}{Evan Patterson}, \bibinfo{person}{Robert
  McBurney}, \bibinfo{person}{Holly Schmidt}, \bibinfo{person}{Ioana Baldini},
  \bibinfo{person}{A Mojsilovi{\'c}}, {and} \bibinfo{person}{Kush~R Varshney}.}
  \bibinfo{year}{2017}\natexlab{}.
\newblock \showarticletitle{Dataflow representation of data analyses: Toward a
  platform for collaborative data science}.
\newblock \bibinfo{journal}{\emph{IBM Journal of Research and Development}}
  \bibinfo{volume}{61}, \bibinfo{number}{6} (\bibinfo{year}{2017}),
  \bibinfo{pages}{9--1}.
\newblock


\bibitem[\protect\citeauthoryear{Pawlowski and Robey}{Pawlowski and
  Robey}{2004}]%
        {pawlowski2004bridging}
\bibfield{author}{\bibinfo{person}{Suzanne~D Pawlowski} {and}
  \bibinfo{person}{Daniel Robey}.} \bibinfo{year}{2004}\natexlab{}.
\newblock \showarticletitle{Bridging user organizations: Knowledge brokering
  and the work of information technology professionals}.
\newblock \bibinfo{journal}{\emph{MIS quarterly}} (\bibinfo{year}{2004}),
  \bibinfo{pages}{645--672}.
\newblock


\bibitem[\protect\citeauthoryear{Peng}{Peng}{2015}]%
        {peng2015reproducibility}
\bibfield{author}{\bibinfo{person}{Roger Peng}.}
  \bibinfo{year}{2015}\natexlab{}.
\newblock \showarticletitle{The reproducibility crisis in science: A
  statistical counterattack}.
\newblock \bibinfo{journal}{\emph{Significance}} \bibinfo{volume}{12},
  \bibinfo{number}{3} (\bibinfo{year}{2015}), \bibinfo{pages}{30--32}.
\newblock


\bibitem[\protect\citeauthoryear{ProACT}{ProACT}{2015}]%
        {ProACT}
\bibfield{author}{\bibinfo{person}{ProACT}.} \bibinfo{year}{2015}\natexlab{}.
\newblock \bibinfo{title}{The DREAM Phil Bowen ALS Prediction Prize4Life
  Challenge, The DREAM ALS Stratification Prize4Life Challenge}.
\newblock
\newblock
\urldef\tempurl%
\url{https://nctu.partners.org/ProACT/Document/DisplayLatest/3}
\showURL{%
\tempurl}


\bibitem[\protect\citeauthoryear{Ribes and Lee}{Ribes and Lee}{2010}]%
        {ribes2010sociotechnical}
\bibfield{author}{\bibinfo{person}{David Ribes} {and}
  \bibinfo{person}{Charlotte~P Lee}.} \bibinfo{year}{2010}\natexlab{}.
\newblock \showarticletitle{Sociotechnical studies of cyberinfrastructure and
  e-research: Current themes and future trajectories}.
\newblock \bibinfo{journal}{\emph{Computer Supported Cooperative Work (CSCW)}}
  \bibinfo{volume}{19}, \bibinfo{number}{3-4} (\bibinfo{year}{2010}),
  \bibinfo{pages}{231--244}.
\newblock


\bibitem[\protect\citeauthoryear{Rolland and Lee}{Rolland and Lee}{2013}]%
        {rolland2013beyond}
\bibfield{author}{\bibinfo{person}{Betsy Rolland} {and}
  \bibinfo{person}{Charlotte~P Lee}.} \bibinfo{year}{2013}\natexlab{}.
\newblock \showarticletitle{Beyond trust and reliability: reusing data in
  collaborative cancer epidemiology research}. In
  \bibinfo{booktitle}{\emph{Proceedings of the 2013 conference on Computer
  supported cooperative work}}. ACM, \bibinfo{pages}{435--444}.
\newblock


\bibitem[\protect\citeauthoryear{Rule, Tabard, and Hollan}{Rule
  et~al\mbox{.}}{2018}]%
        {rule2018exploration}
\bibfield{author}{\bibinfo{person}{Adam Rule}, \bibinfo{person}{Aur{\'e}lien
  Tabard}, {and} \bibinfo{person}{James~D Hollan}.}
  \bibinfo{year}{2018}\natexlab{}.
\newblock \showarticletitle{Exploration and explanation in computational
  notebooks}. In \bibinfo{booktitle}{\emph{Proceedings of the 2018 CHI
  Conference on Human Factors in Computing Systems}}. ACM, \bibinfo{pages}{32}.
\newblock


\bibitem[\protect\citeauthoryear{Schroeder}{Schroeder}{2007}]%
        {schroeder2007research}
\bibfield{author}{\bibinfo{person}{Ralph Schroeder}.}
  \bibinfo{year}{2007}\natexlab{}.
\newblock \showarticletitle{e-Research Infrastructures and Open Science:
  Towards a New System of Knowledge Production?}
\newblock \bibinfo{journal}{\emph{Prometheus}} \bibinfo{volume}{25},
  \bibinfo{number}{1} (\bibinfo{year}{2007}), \bibinfo{pages}{1--17}.
\newblock


\bibitem[\protect\citeauthoryear{Shamekhi, Liao, Wang, Bellamy, and
  Erickson}{Shamekhi et~al\mbox{.}}{2018}]%
        {shamekhi2018face}
\bibfield{author}{\bibinfo{person}{Ameneh Shamekhi}, \bibinfo{person}{Q~Vera
  Liao}, \bibinfo{person}{Dakuo Wang}, \bibinfo{person}{Rachel~KE Bellamy},
  {and} \bibinfo{person}{Thomas Erickson}.} \bibinfo{year}{2018}\natexlab{}.
\newblock \showarticletitle{Face Value? Exploring the effects of embodiment for
  a group facilitation agent}. In \bibinfo{booktitle}{\emph{Proceedings of the
  2018 CHI Conference on Human Factors in Computing Systems}}. ACM,
  \bibinfo{pages}{391}.
\newblock


\bibitem[\protect\citeauthoryear{Sijtsma}{Sijtsma}{2016}]%
        {sijtsma2016playing}
\bibfield{author}{\bibinfo{person}{Klaas Sijtsma}.}
  \bibinfo{year}{2016}\natexlab{}.
\newblock \showarticletitle{Playing with data or how to discourage questionable
  research practices and stimulate researchers to do things right}.
\newblock \bibinfo{journal}{\emph{Psychometrika}} \bibinfo{volume}{81},
  \bibinfo{number}{1} (\bibinfo{year}{2016}), \bibinfo{pages}{1--15}.
\newblock


\bibitem[\protect\citeauthoryear{Spencer~Jr, Butler, Ricker, Marcusiu, Finholt,
  Foster, Kesselman, and Birnholtz}{Spencer~Jr et~al\mbox{.}}{2008}]%
        {spencer200818}
\bibfield{author}{\bibinfo{person}{BF Spencer~Jr}, \bibinfo{person}{Randal
  Butler}, \bibinfo{person}{Kathleen Ricker}, \bibinfo{person}{Doru Marcusiu},
  \bibinfo{person}{Thomas~A Finholt}, \bibinfo{person}{Ian Foster},
  \bibinfo{person}{Carl Kesselman}, {and} \bibinfo{person}{Jeremy~P
  Birnholtz}.} \bibinfo{year}{2008}\natexlab{}.
\newblock \showarticletitle{18 NEESgrid: Lessons Learned for Future
  Cyberinfrastructure Development}.
\newblock \bibinfo{journal}{\emph{Scientific Collaboration on the Internet}}
  (\bibinfo{year}{2008}), \bibinfo{pages}{331}.
\newblock


\bibitem[\protect\citeauthoryear{Staddon}{Staddon}{2017}]%
        {staddon2017scientific}
\bibfield{author}{\bibinfo{person}{John Staddon}.}
  \bibinfo{year}{2017}\natexlab{}.
\newblock \bibinfo{booktitle}{\emph{Scientific Method: How Science Works, Fails
  to Work, and Pretends to Work}}.
\newblock \bibinfo{publisher}{Routledge}.
\newblock


\bibitem[\protect\citeauthoryear{Stasser and Titus}{Stasser and Titus}{1985}]%
        {stasser1985pooling}
\bibfield{author}{\bibinfo{person}{Garold Stasser} {and}
  \bibinfo{person}{William Titus}.} \bibinfo{year}{1985}\natexlab{}.
\newblock \showarticletitle{Pooling of unshared information in group decision
  making: Biased information sampling during discussion.}
\newblock \bibinfo{journal}{\emph{Journal of personality and social
  psychology}} \bibinfo{volume}{48}, \bibinfo{number}{6}
  (\bibinfo{year}{1985}), \bibinfo{pages}{1467}.
\newblock


\bibitem[\protect\citeauthoryear{Stewart and Stasser}{Stewart and
  Stasser}{1998}]%
        {stewart1998sampling}
\bibfield{author}{\bibinfo{person}{Dennis~D Stewart} {and}
  \bibinfo{person}{Garold Stasser}.} \bibinfo{year}{1998}\natexlab{}.
\newblock \showarticletitle{The sampling of critical, unshared information in
  decision-making groups: the role of an informed minority}.
\newblock \bibinfo{journal}{\emph{European Journal of Social Psychology}}
  \bibinfo{volume}{28}, \bibinfo{number}{1} (\bibinfo{year}{1998}),
  \bibinfo{pages}{95--113}.
\newblock


\bibitem[\protect\citeauthoryear{Tan, Wang, and Sabanovic}{Tan
  et~al\mbox{.}}{2018}]%
        {tan2018projecting}
\bibfield{author}{\bibinfo{person}{Haodan Tan}, \bibinfo{person}{Dakuo Wang},
  {and} \bibinfo{person}{Selma Sabanovic}.} \bibinfo{year}{2018}\natexlab{}.
\newblock \showarticletitle{Projecting Life Onto Robots: The Effects of
  Cultural Factors and Design Type on Multi-Level Evaluations of Robot
  Anthropomorphism}. In \bibinfo{booktitle}{\emph{2018 27th IEEE International
  Symposium on Robot and Human Interactive Communication (RO-MAN)}}. IEEE,
  \bibinfo{pages}{129--136}.
\newblock


\bibitem[\protect\citeauthoryear{Thackara}{Thackara}{2000}]%
        {thackara2000edge}
\bibfield{author}{\bibinfo{person}{John Thackara}.}
  \bibinfo{year}{2000}\natexlab{}.
\newblock \showarticletitle{Edge effects: the design challenge of the pervasive
  interface}. In \bibinfo{booktitle}{\emph{CHI'00 Extended Abstracts on Human
  Factors in Computing Systems}}. ACM, \bibinfo{pages}{199--200}.
\newblock


\bibitem[\protect\citeauthoryear{Trifacta}{Trifacta}{2019}]%
        {Trifacta}
\bibfield{author}{\bibinfo{person}{Trifacta}.} \bibinfo{year}{2019}\natexlab{}.
\newblock \bibinfo{title}{Trifacta}.
\newblock
\newblock
\urldef\tempurl%
\url{https://www.trifacta.com/}
\showURL{%
\tempurl}


\bibitem[\protect\citeauthoryear{Tversky and Kahneman}{Tversky and
  Kahneman}{1974}]%
        {tversky1974judgment}
\bibfield{author}{\bibinfo{person}{Amos Tversky} {and} \bibinfo{person}{Daniel
  Kahneman}.} \bibinfo{year}{1974}\natexlab{}.
\newblock \showarticletitle{Judgment under uncertainty: Heuristics and biases}.
\newblock \bibinfo{journal}{\emph{Science}} \bibinfo{volume}{185},
  \bibinfo{number}{4157} (\bibinfo{year}{1974}), \bibinfo{pages}{1124--1131}.
\newblock


\bibitem[\protect\citeauthoryear{Van~Knippenberg and Schippers}{Van~Knippenberg
  and Schippers}{2007}]%
        {van2007work}
\bibfield{author}{\bibinfo{person}{Daan Van~Knippenberg} {and}
  \bibinfo{person}{Michaela~C Schippers}.} \bibinfo{year}{2007}\natexlab{}.
\newblock \showarticletitle{Work group diversity}.
\newblock \bibinfo{journal}{\emph{Annual review of psychology}}
  \bibinfo{volume}{58} (\bibinfo{year}{2007}).
\newblock


\bibitem[\protect\citeauthoryear{Vanschoren, Van~Rijn, Bischl, and
  Torgo}{Vanschoren et~al\mbox{.}}{2014}]%
        {vanschorenOpenMLNetworkedScience2014}
\bibfield{author}{\bibinfo{person}{Joaquin Vanschoren}, \bibinfo{person}{Jan~N.
  Van~Rijn}, \bibinfo{person}{Bernd Bischl}, {and} \bibinfo{person}{Luis
  Torgo}.} \bibinfo{year}{2014}\natexlab{}.
\newblock \showarticletitle{{{OpenML}}: Networked Science in Machine Learning}.
\newblock \bibinfo{journal}{\emph{ACM SIGKDD Explorations Newsletter}}
  \bibinfo{volume}{15}, \bibinfo{number}{2} (\bibinfo{year}{2014}),
  \bibinfo{pages}{49--60}.
\newblock


\bibitem[\protect\citeauthoryear{Velden}{Velden}{2013}]%
        {veldenExplainingFieldDifferences2013}
\bibfield{author}{\bibinfo{person}{Theresa Velden}.}
  \bibinfo{year}{2013}\natexlab{}.
\newblock \showarticletitle{Explaining Field Differences in Openness and
  Sharing in Scientific Communities}. In \bibinfo{booktitle}{\emph{Proceedings
  of the 2013 Conference on {{Computer}} Supported Cooperative Work}}.
  \bibinfo{publisher}{{ACM}}, \bibinfo{pages}{445--458}.
\newblock


\bibitem[\protect\citeauthoryear{Vicente-S{\'a}ez and
  Mart{\'\i}nez-Fuentes}{Vicente-S{\'a}ez and Mart{\'\i}nez-Fuentes}{2018}]%
        {vicente2018open}
\bibfield{author}{\bibinfo{person}{Rub{\'e}n Vicente-S{\'a}ez} {and}
  \bibinfo{person}{Clara Mart{\'\i}nez-Fuentes}.}
  \bibinfo{year}{2018}\natexlab{}.
\newblock \showarticletitle{Open Science now: A systematic literature review
  for an integrated definition}.
\newblock \bibinfo{journal}{\emph{Journal of business research}}
  \bibinfo{volume}{88} (\bibinfo{year}{2018}), \bibinfo{pages}{428--436}.
\newblock


\bibitem[\protect\citeauthoryear{Wang}{Wang}{2016}]%
        {wang2016people}
\bibfield{author}{\bibinfo{person}{Dakuo Wang}.}
  \bibinfo{year}{2016}\natexlab{}.
\newblock \showarticletitle{How people write together now: Exploring and
  supporting today's computer-supported collaborative writing}. In
  \bibinfo{booktitle}{\emph{Proceedings of the 19th ACM Conference on Computer
  Supported Cooperative Work and Social Computing Companion}}. ACM,
  \bibinfo{pages}{175--179}.
\newblock


\bibitem[\protect\citeauthoryear{Wang, Olson, Zhang, Nguyen, and Olson}{Wang
  et~al\mbox{.}}{2015}]%
        {wang2015docuviz}
\bibfield{author}{\bibinfo{person}{Dakuo Wang}, \bibinfo{person}{Judith~S
  Olson}, \bibinfo{person}{Jingwen Zhang}, \bibinfo{person}{Trung Nguyen},
  {and} \bibinfo{person}{Gary~M Olson}.} \bibinfo{year}{2015}\natexlab{}.
\newblock \showarticletitle{DocuViz: visualizing collaborative writing}. In
  \bibinfo{booktitle}{\emph{Proceedings of the 33rd Annual ACM Conference on
  Human Factors in Computing Systems}}. ACM, \bibinfo{pages}{1865--1874}.
\newblock


\bibitem[\protect\citeauthoryear{Wang, Tan, and Lu}{Wang et~al\mbox{.}}{2017}]%
        {wang2017users}
\bibfield{author}{\bibinfo{person}{Dakuo Wang}, \bibinfo{person}{Haodan Tan},
  {and} \bibinfo{person}{Tun Lu}.} \bibinfo{year}{2017}\natexlab{}.
\newblock \showarticletitle{Why users do not want to write together when they
  are writing together: Users' rationales for today's collaborative writing
  practices}.
\newblock \bibinfo{journal}{\emph{Proceedings of the ACM on Human-Computer
  Interaction}} \bibinfo{volume}{1}, \bibinfo{number}{CSCW}
  (\bibinfo{year}{2017}), \bibinfo{pages}{107}.
\newblock


\bibitem[\protect\citeauthoryear{Wang, Wang, Yu, Ashktorab, and Tan}{Wang
  et~al\mbox{.}}{2019}]%
        {wang2019slack}
\bibfield{author}{\bibinfo{person}{Dakuo Wang}, \bibinfo{person}{Haoyu Wang},
  \bibinfo{person}{Mo Yu}, \bibinfo{person}{Zahra Ashktorab}, {and}
  \bibinfo{person}{Ming Tan}.} \bibinfo{year}{2019}\natexlab{}.
\newblock \showarticletitle{Slack Channels Ecology in Enterprises: How
  Employees Collaborate Through Group Chat}.
\newblock \bibinfo{journal}{\emph{arXiv preprint arXiv:1906.01756}}
  (\bibinfo{year}{2019}).
\newblock


\bibitem[\protect\citeauthoryear{Warr}{Warr}{2006}]%
        {warr2006situated}
\bibfield{author}{\bibinfo{person}{Andrew Warr}.}
  \bibinfo{year}{2006}\natexlab{}.
\newblock \showarticletitle{Situated and distributed design}. In
  \bibinfo{booktitle}{\emph{NordiCHI Workshop on Distributed Participatory
  Design, Oslo, Norway}}. Citeseer.
\newblock


\bibitem[\protect\citeauthoryear{Wenger}{Wenger}{2010}]%
        {wenger2010communities}
\bibfield{author}{\bibinfo{person}{Etienne Wenger}.}
  \bibinfo{year}{2010}\natexlab{}.
\newblock \showarticletitle{Communities of practice and social learning
  systems: the career of a concept}.
\newblock In \bibinfo{booktitle}{\emph{Social learning systems and communities
  of practice}}. \bibinfo{publisher}{Springer}, \bibinfo{pages}{179--198}.
\newblock


\bibitem[\protect\citeauthoryear{Woelfle, Olliaro, and Todd}{Woelfle
  et~al\mbox{.}}{2011}]%
        {woelfle2011open}
\bibfield{author}{\bibinfo{person}{Michael Woelfle}, \bibinfo{person}{Piero
  Olliaro}, {and} \bibinfo{person}{Matthew~H Todd}.}
  \bibinfo{year}{2011}\natexlab{}.
\newblock \showarticletitle{Open science is a research accelerator}.
\newblock \bibinfo{journal}{\emph{Nature Chemistry}} \bibinfo{volume}{3},
  \bibinfo{number}{10} (\bibinfo{year}{2011}), \bibinfo{pages}{745}.
\newblock


\bibitem[\protect\citeauthoryear{Wulf}{Wulf}{1989}]%
        {wulf1989national}
\bibfield{author}{\bibinfo{person}{William~A Wulf}.}
  \bibinfo{year}{1989}\natexlab{}.
\newblock \showarticletitle{The national collaboratory-A white paper}.
\newblock  (\bibinfo{year}{1989}).
\newblock


\bibitem[\protect\citeauthoryear{Zhang}{Zhang}{2018}]%
        {zhang2018jupyterlab_voyager}
\bibfield{author}{\bibinfo{person}{Ji Zhang}.} \bibinfo{year}{2018}\natexlab{}.
\newblock \showarticletitle{JupyterLab\_Voyager: a Data Visualization
  Enhancement in JupyterLab}.
\newblock  (\bibinfo{year}{2018}).
\newblock


\end{thebibliography}

%







\end{document}